\documentclass[preprint,11pt,authoryear]{elsarticle}
\usepackage{iftex}
\ifPDFTeX
  \usepackage[T1]{fontenc}
  \usepackage[utf8]{inputenc}
  \usepackage{tgpagella}
\else
  \usepackage{fontspec}
\fi
\usepackage{fancyhdr}
\usepackage{geometry}
\usepackage{amssymb}
\usepackage{threeparttable}
\usepackage{lipsum}
\usepackage{graphicx} 
\usepackage{subcaption}
\usepackage{amsthm}
\usepackage{pdflscape}
\usepackage{amsmath}
\usepackage{lineno}
\usepackage{hyperref}
\usepackage{tabularx}
\usepackage{svg}
\usepackage{url}
\usepackage{makecell}
\usepackage{cleveref}
\setcitestyle{numbers,square}
\usepackage{bm}
\usepackage{amsmath}
\usepackage{amsfonts}
\usepackage{float}
\usepackage{xr}
\usepackage{framed}
\usepackage{makecell}
\usepackage{arydshln}
\usepackage{fancyhdr}
\usepackage{booktabs}
\usepackage{enumitem}
\usepackage{subcaption}
\usepackage{longtable}
\usepackage{ragged2e} 
\usepackage{array}
\usepackage{tcolorbox}
\usepackage{listings}
\usepackage{algorithm}
\usepackage{algpseudocode}
\newcolumntype{P}[1]{>{\RaggedRight\arraybackslash}p{#1}}
\usepackage{colortbl}
\usepackage{array}
\usepackage{tabularray}
\usepackage{float}
\usepackage{graphicx}
\usepackage{rotating}
\usepackage[normalem]{ulem}
\NewTableCommand{\tinytableDefineColor}[3]{\definecolor{#1}{#2}{#3}}

\usepackage{xcolor}

\usepackage{multirow}

\tcbset{
  promptcardstyle/.style={
    colback=blue!3,
    colframe=blue!40!black,
    fonttitle=\bfseries,
  }
}

\newtcolorbox{promptcard}[1]{
  promptcardstyle,
  title=#1,
}

\newcommand{\sectionname}{Section}
\usepackage{xcolor}

\usepackage{lineno}
\journal{arXiv.org}
\hypersetup{
    pdfauthor={RERITE},
    pdftitle={Measuring the State of Open Science in Transportation Using Large Language Models}
}
\begin{document}
\begin{frontmatter}
\title{\textbf{\Large Measuring the State of Open Science in Transportation\\Using Large Language Models}}

\author[inst2]{Junyi Ji\fnref{fn1}}
\author[inst1]{Ruth Lu\fnref{fn1}}
\author[inst3]{Linda Belkessa\fnref{fn2}}
\author[inst4]{Liming Wang\fnref{fn2}}
\author[inst5]{Silvia Varotto\fnref{fn2}}
\author[inst6]{Yongqi Dong}
\author[inst7]{Nicolas Saunier}
\author[inst3]{Mostafa Ameli}
\author[inst8]{Gregory S. Macfarlane}
\author[inst5]{Bahman Madadi}
\author[inst1]{Cathy Wu\corref{cor1}}

\cortext[cor1]{Corresponding author: cathywu@mit.edu}

\affiliation[inst1]{organization={Massachusetts Institute of Technology},
location={Boston, MA},
country={United States}}

\affiliation[inst2]{organization={Vanderbilt University},
location={Nashville, TN},
country={United States}}

\affiliation[inst3]{organization={Université Gustave Eiffel},
location={Paris},
country={France}}

\affiliation[inst4]{organization={Portland State University},
location={Portland, OR},
country={United States}}

\affiliation[inst5]{organization={École Nationale des Travaux Publics de l'État (ENTPE)},
location={Vaulx-en-Velin},
country={France}}

\affiliation[inst6]{organization={Delft University of Technology},
location={Delft},
country={The Netherlands}}

\affiliation[inst7]{organization={Polytechnique Montréal},
location={Montréal},
country={Canada}}

\affiliation[inst8]{organization={Brigham Young University},
location={Provo, UT},
country={United States}}

\fntext[fn1]{Equal contribution}
\fntext[fn2]{Equal contribution}

\begin{abstract}
Open science initiatives have strengthened scientific integrity and accelerated research progress across many fields, but the state of their practice within transportation research remains under-investigated. Key features of open science, defined here as data and code availability, are difficult to extract due to the inherent complexity of the field. Previous work has either been limited to small-scale studies due to the labor-intensive nature of manual analysis or has relied on large-scale bibliometric approaches that sacrifice contextual richness.
This paper introduces an automatic and scalable feature-extraction pipeline that combines full-text extraction and human agreement analysis in order to measure code and data availability in transportation research. We employ Large Language Models (LLMs) for this task and validate their performance against a manually curated dataset and through an inter-rater agreement analysis. We applied this pipeline to examine 10,724 research articles published in the Transportation Research Part series of journals between 2019 and 2024.
Our analysis found that only 5\% of quantitative papers shared a code repository, 4\% of quantitative papers shared a data repository, and about 3\% of papers shared both,
with trends differing across journals, topics, and geographic regions. We found no significant difference in citation counts or review duration between papers that provided data and code and those that did not, suggesting a misalignment between open science efforts and traditional academic metrics. Consequently, encouraging these practices will likely require structural interventions from journals and funding agencies to supplement the lack of direct author incentives. The pipeline developed in this study can be readily scaled to other journals, representing a critical step toward the automated measurement and monitoring of open science practices in transportation research. An interactive explorer can be found on the project page \url{https://www.rerite.org/MOST}.
\end{abstract}

\end{frontmatter}

\section{Introduction}
\label{sec:intro}
\subsection{Motivation and vision}
Open science initiatives have accelerated research progress across many disciplines, including computational science~\cite{peng2011reproducible}, psychology~\cite{open2015estimating}, and statistics~\cite{stodden2014reproducible}. Recently, open science practices have gained increasing recognition in the transportation research community, as reflected by the introduction of data availability statements in Transportation Research journals~\cite{trc_guide2025}. Notably, \textit{Transportation Research Part C} explicitly emphasizes and promotes open science in its aims and scope, stating that ``Special emphasis is given in open science initiatives and promoting the opening of large-scale datasets that can support transferability and benchmarking of different approaches~\cite{trc_aimsscope2025}.'' However, despite these policy advancements, the extent to which open science practices have been adopted and implemented in transportation research remains largely unexamined.

The idea of open science \cite{woelfle2011open} is to make scientific knowledge \textit{openly available, accessible, and reusable}, as defined by the United Nations Educational, Scientific and Cultural Organization (UNESCO)~\cite{unesco_openscience2022}. As open science has gained awareness~\cite{awinker2023equity,gentemann2023nasa}, open science monitoring (OSM) has emerged as a research direction and has been endorsed by UNESCO. OSM is not merely bookkeeping: transparent, responsible measurement is a practical mechanism for guiding, evaluating, and accelerating the transformation of research practices. In 2025, core principles for OSM were established~\cite{bobrov_2025_17853227}, emphasizing (\romannumeral1) relevance and significance, (\romannumeral2) transparency and reproducibility, and (\romannumeral3) self-assessment and responsible use.

Open science encompasses the ideas of reproducibility and replicability in science \cite{national2019reproducibility}, where reproducibility refers to the ability to obtain the same results using the same data and methods, while replicability refers to the ability to obtain similar results using different data or methods. Making data and code available is the first and essential step towards achieving reproducibility and replicability~\cite{cruwell2019seven}, as it allows other researchers to verify and build upon the original work.  
A recent study within the artificial intelligence (AI) research community suggests that for the purpose of successful replication, the quality of shared code is a secondary consideration to its mere availability~\cite{gundersen2025unreasonable}. 
This highlights the essential role of open code and data in promoting both reproducibility and replicability and motivates a systematic measurement of open science practices in transportation research, with a particular focus on the availability of data and code in papers published in Transportation Research journals. These motivate us to build a repeatable, low-cost system for monitoring open code and data practices in transportation research, helping the field make steady, evidence-based progress toward broader open science adoption.

\subsection{Challenges}
Measuring the state of open science practices is a complex and resource-intensive endeavor~\cite{stagge2019assessing,yang2020estimating,youyou2023discipline}. Traditionally, this involves manually reviewing each paper to assess the availability of data and code~\cite{stagge2019assessing}, as well as verifying the validity of provided resources. Such manual processes are labor-intensive, time-consuming~\cite{stagge2019assessing,pineau2021improving,seibold2021computational,Olszewski2023}, and do not scale efficiently to large corpora of research articles. And without proper training and clear instructions, manual annotation, especially when involving multiple annotators, can easily produce inconsistent and inaccurate results. A recent study~\cite{riehl2025revisiting} developed automated approaches based on keyword and link extraction using regular expressions to improve scalability. The proposed pipeline in~\cite{riehl2025revisiting} is reproducible and extensible, and we therefore benchmark it in this work as well. However, the regular expressions often fail to capture the surrounding context and thus generate false positives, for example, detecting benchmark repositories or GitHub links in citations as evidence of sharing. The trade-offs between accuracy and scalability are summarized in \tablename~\ref{tab:tradeoff}.

\begin{table}[htbp]
\centering
\caption{Accuracy and scalability trade-offs for code and data availability extraction from publications. We analyze LLM accuracy relative to human annotators in Section~\ref{sec:agreement-analysis-results} through a multi-rater agreement analysis, which further confirms LLM capability for such feature extraction.}
\label{tab:tradeoff}
\small
\begin{tabular}{lcc}
\toprule
            & Slow and labor-intensive & Fast and scalable \\
\hline
Accurate    & Manual annotation (clear guidelines) & \colorbox{green!20}{LLMs (this study)} \\
Inaccurate  &  Manual annotation (with ambiguity) & Regular expressions \\
\bottomrule
\end{tabular}
\end{table}

Furthermore, transportation research poses unique challenges for measuring open science practices because “data and code availability” is often conceptually under-specified and therefore difficult to operationalize. In particular, the boundary between using publicly accessible resources and making research artifacts available is frequently blurred. For example, many studies rely on widely used public datasets such as NGSIM~\cite{NGSIM}, yet do not release the specific processed or derived datasets they create for tasks such as calibrating car-following or lane-changing models. In such cases, the study benefits from open data, but the research-ready data needed to replicate the analysis may not be available. Conversely, even when raw data are proprietary or sensitive, researchers can often share post-processed or aggregated derivatives that enable partial or substantial reuse. These scenarios raise a measurement dilemma: should a paper be credited for “using publicly available data,” for “making data available,” for both, or for neither? This ambiguity implies that data and code availability in transportation research is not binary, but spans a spectrum from minimal reliance on existing open resources to proactive dissemination of reusable artifacts that sustain open science. These definitions need to be formalized upfront to ensure transparency and reproducibility before any measurement is conducted, regardless of the method used.

\subsection{Problem statement and contributions}
To address these challenges, this study first formalizes the operational definitions and decision rules required to distinguish between using or citing an existing public dataset, releasing processed research data, and releasing research code.  We then introduce an automated framework that balances accuracy with scalability to measure open science practices in transportation research. Focusing on code and data availability as observable artifacts in full-text articles, we develop an extraction pipeline that using large language models (LLMs) to capture contextual signals that keyword searches and regular expressions can miss. The pipeline processes Elsevier \texttt{XML} files at a scale of over 10,000 papers, identifies and validates artifact links, classifies availability, and is evaluated against a human validation dataset.

Using these large-scale extracted features, we address the following research questions:
\begin{enumerate}[label=(Q\arabic*),noitemsep, parsep=1pt, topsep=1pt, partopsep=0pt]
\item What is the proportion of papers that make code and data available?
\item Which factors are associated with the availability of code and data? 
\item Is code and data availability simply a matter of time, i.e., to what extent is open science behavior incentivized, or are external policies warranted?
\end{enumerate}

The broader vision of this study is to establish a framework for monitoring open science practices in transportation research. Rather than emphasizing current gaps, we view the measurement of open science practices as a catalyst which enables the community to identify common challenges, recognize progress, and foster a culture of collaboration and reproducibility that advances scientific rigor and innovation in transportation research.

The contributions of this paper are summarized as follows:
\begin{enumerate}[label=(\roman*),noitemsep, parsep=1pt, topsep=1pt, partopsep=0pt]
\item We develop and implement a scalable pipeline that extracts code and data availability and its contributing factors in full-text Transportation Research articles, enabled by LLMs.
\item We validate all LLM-extracted features with a manually annotated dataset and keep only those that meet established inter-rater agreement standards for downstream analysis.
\item We provide the first large-scale, field-wide analysis of the state of open science practices in transportation research, capturing the effect of temporal, topical, and geographical factors on code and data availability and outlining implications for accelerating open science in the community. An interactive explorer can be found on the project page \url{https://www.rerite.org/MOST}.
\end{enumerate}

The scope of our work focuses on sharing code and data used in research via online repositories. A full glossary of open scholarship terms can be found at \cite{parsons2022community}. There are many other dimensions of open science and reproducibility that are outside the scope of this paper, such as paywalls, institutional access, quality of documentation, computational reproducibility, and more. We hope that these can be further explored in future works.

\subsection{Organization of the paper}
The remainder of the paper is organized as follows. 
\sectionname~\ref{sec:related} situates the study within related work on open science in transportation and open science monitoring. 
\sectionname~\ref{sec:definition} defines the terminology and operational criteria used for identifying data and code availability.  
\sectionname~\ref{sec:data} outlines the full-text dataset curation and the manual annotation procedure. 
\sectionname~\ref{sec:feature-extraction} presents the automatic feature extraction pipeline, including meta-data parsing, topic modeling, LLM-based extraction, agreement analysis, and postprocessing. 
\sectionname~\ref{sec:analysis} reports descriptive findings and estimates choice models to examine factors associated with data and code availability. 
\sectionname~\ref{sec:discussion} discusses the results and performance. 
\sectionname~\ref{sec:recommendations} provides implementation guidance and community-facing recommendations. 
\sectionname~\ref{sec:conclusion} concludes the paper, summarizes limitations, and identifies directions for future work.

\section{Related Work}
\label{sec:related}

\subsection{Open science in transportation research}
\subsubsection{Signs of momentum}
Open science practices, including open-access publishing, open data, and open-source software, support reproducibility by making research materials transparent and accessible. Awareness of open science principles in transportation research, particularly regarding availability, transparency, and reproducibility, emerged in the 2010s. To address the ``data deluge'' in modern transportation research, comply with U.S. federal government policy changes, and maximize returns on research investments, Purdue University Libraries and its Joint Transportation Research Program~\cite{newton2012engaging,zilinski2014evolution} developed workflows for archiving transportation research outputs and data. These workflows covered creation, publication, and dissemination planning. These efforts highlighted a recurring issue: while many researchers were willing to share their data, they often lacked effective mechanisms to do so.
Parallel developments occurred in Europe. Open science became a legal obligation under Horizon Europe (2021-2027)~\cite{nielsen2023using}, the EU’s flagship program for supporting research and innovation. A common challenge is that research data is typically confined to internal use and often becomes unavailable after project completion. To address this, a data sharing framework (DSF) for naturalistic driving studies (NDS)~\cite{gellerman2016data} was developed and demonstrated through the EU \texttt{UDRIVE} project~\cite{eenink2014udrive}, the first EU project to explicitly address data-sharing prerequisites. 

Similar reuse needs exist for code and tools in transportation research. As estimated in~\cite{tamminga2012design}, 30\%–50\% of the total effort required to develop a simulation case study is devoted to data-related tasks and network preparation, including calibration and validation that rely on extensive field data. Compounding this inefficiency, a significant portion of the effort is spent rebuilding basic functionality and repeatedly reinventing the wheel. Making tools and code openly available could therefore lead to substantial collective savings in time and effort. Motivated by this need, \cite{tamminga2012design} subsequently developed \texttt{Open Traffic}~\cite{tamminga2014open}, which provides standardized, multi-scale traffic problems and shared network settings to support researchers worldwide in developing and comparing methods.

More efforts have been conducted in specific transportation sub-fields, for example, car-following model calibration~\cite{punzo2021calibration}, sensor-based mobility mode recognition~\cite{wang2019enabling}, and discrete choice models for travel demand modeling~\cite{wang2024comparing}. Additional community-driven initiatives have also emerged. The Zephyr Foundation, established in 2017, promotes open-source software and data sharing in transportation research, particularly in transportation and land-use decision-making. The International Association of Travel Behaviour Research (IATBR) has similarly advocated for a fully open science framework in which research outputs are freely accessible and reproducible \cite{caicedo2025sharing}. REproducible Research In Transportation Engineering (RERITE), a transportation research community the authors belong to, is also actively promoting open science principles~\cite{wu2024reproducibility}, developing best practices, and fostering community engagement. Note that transportation research is largely data driven, and widely used open datasets such as NGSIM~\cite{NGSIM}, PeMS~\cite{choe2002freeway}, Transportation Networks~\cite{transportationnetworks}, and Swissmetro~\cite{bierlaire2001acceptance} have played a major role in supporting empirical and methodological advances. Progress in adjacent fields like automated driving and computer vision has been strongly enabled by open science practices, where large-scale public datasets including \texttt{nuScenes}~\cite{caesar2020nuscenes} and the Waymo Open Dataset~\cite{sun2020scalability}, Waymo Open Motion Dataset~\cite{ettinger2021large} have each accumulated thousands of citations and set expectations for transparency, benchmarking, and reproducibility. We also see a growing wave of bottom-up open data efforts from our own research community \cite{barmpounakis2020new,seo2020evaluation,gloudemans202324,yabe2024enhancing,yabe2024yjmob100k,schicktanz2025dlr}. To help promote these initiatives, we maintain a curated list at \href{https://www.rerite.org/awesome-rerite}{\texttt{awesome-rerite}}, collecting them to the best of our knowledge.

\subsubsection{Persistent barriers and opportunities}
Despite recent progress, substantial barriers to sharing data and code remain. Transportation research not only faces these general challenges but often amplifies them due to the nature of the data and the ecosystem in which it is produced. 
Transportation research operates under real-world constraints and complexities that make openness nontrivial.
These constraints include privacy protections~\cite{folla2021main,wood2024reproducibility,shi2025synshrp2}, the large size and sensitivity of datasets~\cite{nielsen2023using}, the heterogeneity of data sources~\cite{welch2019big}, the diversity of regional characteristics~\cite{sun2020identifying}, and even geopolitical security concerns~\cite{crampton2015collect,nikander2024threats,zhang2024china},
all of which complicate efforts to make data and code publicly available.
They are further compounded by the need for active collaboration among diverse stakeholders~\cite{nie2025brief}, including academic institutions, government agencies, private companies, and the public. These intertwined constraints often limit the degree to which data and code can be openly shared, even when researchers are willing to do so.

From a research perspective, transportation research is highly interdisciplinary~\cite{sun2017discovering}, spanning fields such as physics, economics, control theory, operations research, social science, psychology, computational science, and environmental science~\cite{Mahmassani2024}.
This diversity brings valuable perspectives but also creates inconsistencies in data formats, modeling approaches, and expectations for transparency and reproducibility.
Researchers from different backgrounds often hold different views on what constitutes shareable data or reusable code, making it difficult to establish consistent open science practices across the field. Meanwhile, when data and code are made openly available, this diversity becomes a strength: researchers can reuse resources across domains, build on one another’s methods, and accelerate collective progress. Ultimately, greater openness can bridge disciplinary divides, connect research with practice~\cite{NSF2437781}, and foster a more collaborative and transparent research culture.

\subsection{Science of science in transportation research}
The measurement of open science practices belongs to the broader field of \emph{science of science} (SciSci), which uses large-scale literature data to characterize the structure and dynamics of research itself and to inform science policy and research management \cite{fortunato2018science}. Data and code availability is one such measurable dimension \cite{liu2023data}. SciSci methods have been applied extensively in transportation to understand the state of the field and its evolution. Bibliometric studies have examined research trends in individual journals \cite{jiang2020bibliometric, antoniou2025overview}, the field as a whole \cite{sun2017discovering, modak2019fifty, sun2020identifying, wandelt2025review, wandelt2025hitchhiker, zhao2026uncovering}, and specific sub-fields \cite{tamakloe2023discovering, ma2022mapping}, while co-authorship and collaboration networks have also been investigated \cite{sun2017coauthorship}. A related line of work analyzes the review process itself, including the reasons papers are rejected \cite{wu2024your} via 5,000 rejected paper and the corresponding review comments and the governance of peer review in transportation research \cite{bae2026peer}.

A common feature of these studies is their reliance on \emph{structured} signals, for example titles, abstracts, keywords, author affiliations, and citations, that are readily extracted from bibliographic metadata. Far less attention has been paid to the \emph{unstructured} full text of articles, which is precisely where evidence of open science practices, such as statements of code and data availability, typically resides. Our work targets this gap by extracting and analyzing these unstructured signals at scale from full text paper.

\subsection{Monitoring and measuring open science practices}

From a monitoring perspective, measurement of data and code availability falls into two regimes, depending on how journals and venues capture this information:
\begin{itemize}[noitemsep, parsep=1pt, topsep=1pt, partopsep=0pt]
\item Unstructured: publications may mention data and code availability, but do so in free text and inconsistent locations (e.g., acknowledgments, footnotes, methods, supplementary material notes, or informal ``data available upon request'' language). In this regime, reliable monitoring requires robust extraction methods from full-text documents.
\item Structured: publications provide standardized, machine-readable metadata (e.g., required fields or standardized statements) for data and code availability. In this regime, monitoring is comparatively straightforward: extraction is largely a matter of parsing consistent metadata, and extensive post-processing is unnecessary.
\end{itemize}

At present, transportation research largely remains in the unstructured regime. As a result, the central challenge is not just defining what availability means, but also reliably extracting availability features at scale from full-text publications. This study therefore focuses on the unstructured case, while also motivating the structured case as a desirable endpoint: if structured data and code metadata became standard practice, the need for complex extraction tools would diminish substantially. We return to this distinction in the discussion of practices and future monitoring directions.

\subsubsection{Unstructured}
In many fields, measurement efforts start out unstructured, and clarity in definitions determines what should be measured. In signal processing, \cite{vandewalle2009reproducible} proposed a six-level reproducibility scale, ranging from complete irreproducibility to effortless reproduction within 15 min using standard tools. More broadly, \cite{goodman2016does} note that the language and conceptual framework of ``research reproducibility'' remain nonstandard and unsettled across the sciences. To avoid misunderstandings that arise when reproducibility terminology is treated as a proxy for ``truth,'' they consolidate prior taxonomies into three categories: methods reproducibility, results reproducibility, and inferential reproducibility. At a more operational level, reproducibility assessment can be staged from basic to stringent checks: whether code and data are available, whether the code runs, whether it runs with minor fixes, and whether it reproduces the reported results. Additional signals of artifact quality, such as README completeness and documentation, can further inform these assessments.
In transportation research, \cite{ZHENG2021100004} proposed three requirements for reproducible publications: (\romannumeral1) sharing the scripts/codes for generating the entire paper (data analysis, figures, tables, models) using open-source programming languages, (\romannumeral2) openly releasing the data, and (\romannumeral3) clearly documenting the software environment.

Existing evaluation efforts are labor-intensive and therefore difficult to sustain beyond small scales. At the same time, simple pattern-matching approaches \cite{riehl2025revisiting} cannot reliably operationalize detailed definitions of reproducibility or extract the required features. This creates a dilemma between scale and accuracy: manual assessments do not scale, while automated keyword-based methods lack fidelity. As a result, neither approach consistently satisfies key open science monitoring principles, particularly transparency, reproducibility, and responsible use.

Emerging technologies, particularly large language models (LLMs), have demonstrated strong effectiveness in processing and analyzing complex textual data~\cite{gilardi2023chatgpt}, creating new opportunities for open science monitoring. For example, \texttt{Reproscreener} \cite{bhaskar} illustrates how LLMs can support post-publication monitoring by automating checks of machine learning papers for the presence of information and documentation needed to enable replication. However, the presence of such reporting elements does not necessarily guarantee computational reproducibility.
Building on this line of work, \cite{siegelcore} introduced the Computational Reproducibility Agent Benchmark (\texttt{CORE-Bench}), a benchmark suite designed to evaluate LLMs’ capacity to perform end-to-end computational reproducibility checks. Beyond assessing whether reproducibility-related criteria are met, \texttt{CORE-Bench} also involves executing the provided code and data to verify whether computational results can be reproduced. The authors further use LLM embeddings to predict reproducibility scores, offering a computationally efficient alternative to labor-intensive ma\-nual verification. Together, these approaches demonstrate how LLMs can enable scalable monitoring of reproducibility-related practices within the broader open science ecosystem. More broadly, LLMs have been applied to adjacent scholarly workflows relevant to open science monitoring, including peer review automation \cite{lee2025role}, abstract screening \cite{huotala2024promise}, and systematic review writing \cite{mahmoudi2024critical}. Recent work has also explored using LLMs to extract funding information and associated targeted research questions from text at scale \cite{lai2025comparative}. While these applications do not directly operationalize large-scale open science monitoring, they provide scalable components that can be integrated into monitoring systems, and their use for this purpose remains an early but promising direction.

\subsubsection{Structured}
In journals and peer-reviewed conferences such as Nature and its flagship journals, as well as leading AI and data science venues including NeurIPS (Neural Information Processing Systems), ICLR (International Conference on Learning Representations), ICML (International Conference on Machine Learning), and KDD (Knowledge Discovery in Databases), data and code availability are reported in dedicated, structured sections published alongside the main content. Measuring the data and code availability can be done just by reviewing the availability statement alone. One key takeaway is that introducing a structured statement into the submission, review, and camera-ready checks significantly improves code and data availability rates~\cite{hutson2018artificial,pineau2021improving}.

Moreover, the machine learning community is developing metadata frameworks, such as \texttt{Croissant}~\cite{akhtar2024croissant}, to make datasets more readable for both machines and humans. If open science practices become widely accepted in transportation research, availability features could be measured through structured data collected via self-reporting and peer review as part of the publication process. At that stage, this section could serve as a reference for the community to develop the structured checklist.

\subsection{Large Language Models for feature extraction}
As LLMs have grown more powerful, Natural Language Processing (NLP) benchmarks have also become more complex, allowing a better understanding of the possibilities and limitations of LLMs. Many works have assessed the ability of LLMs to recall, extract, and reason from information.

\subsubsection{LLM capability testing on scientific papers}

Papers often require inputs of tens of thousands of tokens, and relevant pieces of information may be scattered across the work. Because of the computational cost of processing more tokens, LLMs have a limited context window. But as the number of parameters has increased, LLMs have been able to handle in longer contexts \cite{hsieh2024ruler}, allowing them to be used for more tasks, such as reading scientific papers.

LLMs have been used in many ways to process papers, from performing literature reviews \cite{agarwal2025litllmtoolkitscientificliterature}, answering factual questions about a work \cite{dasigi2021datasetinformationseekingquestionsanswers} \cite{chen2026rpcbenchfinegrainedbenchmarkresearch}\cite{javaji2025aivalidatesciencebenchmarking}, and even generating follow-up studies \cite{liu2026researchbenchbenchmarkingllmsscientific}. Despite these advances, LLMs still face limitations. Long contexts still damage the question answering ability of LLMs \cite{du2025contextlengthhurtsllm}, and LLM capability works sometimes use an LLM-as-a-judge framework, where a more powerful model is used as ground truth \cite{chen2026rpcbenchfinegrainedbenchmarkresearch}. Because our full dataset is too large to be human-validated, we instead compared the LLM's performance on a small, human-validated sample of the whole set.

\subsubsection{Prompting strategies}

The performance of LLMs can be highly dependent on how the prompt is structured. Various techniques have been used to increase accuracy. Some of the most common include giving examples \cite{zhong2023chatgptunderstandtoocomparative}, asking LLMs to first restate relevant information from long contexts \cite{du2025contextlengthhurtsllm}, inducing chain of thought reasoning \cite{zhong2023chatgptunderstandtoocomparative}, or splitting long tasks into many smaller tasks \cite{zhong2023chatgptunderstandtoocomparative}. Our result on the full dataset uses prompts that split the long tasks into smaller tasks, and we also evaluate the performance of other prompting methods on the small, manual validation dataset.

\section{Definitions and terminology}
\label{sec:definition}
Open science represents a comprehensive vision integrating a wide array of concepts~\cite{vicente2018open, parsons2022community}. However, using such broad terminology, which encompasses concepts ranging from open access publishing to general knowledge sharing, introduces unnecessary ambiguity to this study and the associated feature extraction. Therefore, we narrow the scope of \textit{open science practices} to focus specifically on data and code availability connected with a paper. To align with established open science monitoring principles~\cite{bobrov_2025_17853227}, we adopt clear, transparent decision rules as detailed below.

For data availability, we anchor our criteria to Elsevier's Research Data Guidelines for transportation research journals~\cite{elsevier_research_data_guidelines_2025}:

\begin{quote} 
``You are encouraged or required to: 
\begin{itemize} 
\item Deposit your research data in a relevant data repository; 
\item Cite and link to this dataset in your article; 
\item If this is not possible, make a statement explaining why research data cannot be shared.'' 
\end{itemize} 
\end{quote}

Guided by these guidelines, we identify two distinct data availability features: (\romannumeral1) whether the dataset is deposited in an accessible repository, and (\romannumeral2) whether the dataset is cited (and linked) in the paper. These features capture two different behaviors: depositing data represents a contribution to open science, whereas citing existing datasets indicates benefiting from it. We define "available data" as a publicly deposited dataset that includes the processed inputs. If neither depositing nor citing criteria are met, the paper is classified as having no data available.

Regarding code, while Elsevier guidelines broadly categorize it under research data, we analyze it as a distinct entity. We define code availability as the presence of a public repository or archival deposit containing source files. 

Our scope in this study is restricted to artifact availability detectable within the full text and its links; we do not perform external web searches for resources not explicitly mentioned in the paper. Consequently, data and code availability, as measured here, serves as an early indicator and reasonable proxy for computational reproducibility, rather than a guarantee of full reproducibility~\cite{ZHENG2021100004,seibold2021computational}.

\begin{figure}[htbp]
    \centering
    \includegraphics[width=0.65\linewidth]{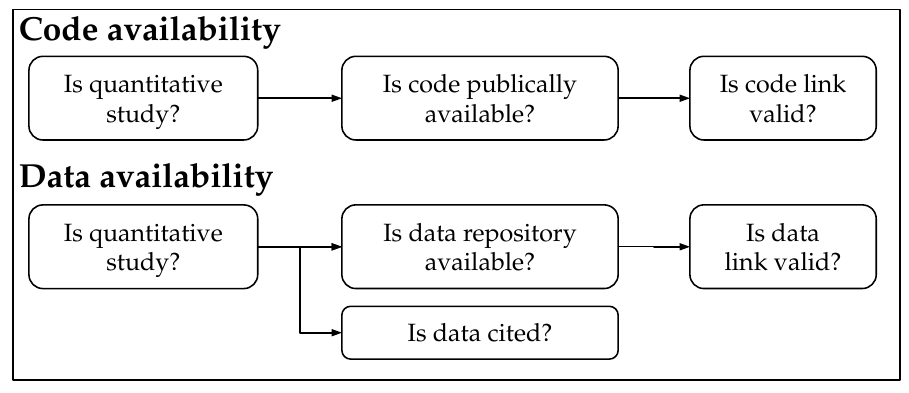}
    \caption{Features considered for data and code availabilities.}
    \label{fig:MVDv2}
\end{figure}

\figurename~\ref{fig:MVDv2} presents the logic used in this paper for determining data and code availability. Conceptually, we avoid counting studies that do not rely on code or data, since including them would underestimate the actual rate of code and data availability. The full definitions of all features shown in the flowchart are provided below.

\subsection{Data availability features}
\begin{enumerate}[label=(\roman*),noitemsep, parsep=1pt, topsep=1pt, partopsep=0pt]

\item Data repository. - Does the paper share a data repository? The repository must be created by the authors, but may contain data collected by others (i.e., processed data from NGSIM).
\item Data citation or link. - Does the paper use and cite or link at least one publicly available dataset?
\item Data repository validity. - Is the repository actually live and does it contain data?
\end{enumerate}

Based on the above definitions, we categorize data availability into four categories:%
\begin{enumerate}[label=(\roman*),noitemsep, parsep=1pt, topsep=1pt, partopsep=0pt]
    \item The first category (NC, NR) represents papers that neither cite data nor provide a data repository. 
    \item The second category (NC, R) includes papers that provide a link to a data repository but do not cite any datasets. These papers may be fully reproducible if they may collect and use original data. 
    \item The third category (C, NR) consists of papers that cite datasets but do not provide a data repository. 
    \item The fourth category (C, R) encompasses papers that both cite datasets and provide a data repository. 
\end{enumerate}
This taxonomy allows for a nuanced understanding of data availability practices in transportation research.

\subsection{Code availability features}
\begin{enumerate}[label=(\roman*),noitemsep, parsep=1pt, topsep=1pt, partopsep=0pt]
\item Code dependence. (``Is quantitative study?'') - Did the paper do any quantitative calculations?
\item Code availability. - Does the paper share a link to code used in the text of the paper?
\item Code deposit link validity - Is the link actually live and does it contain code?
\end{enumerate}
Similar to data availability, we categorize code availability into two categories: code available (CA) and code unavailable (CU).

We use these definitions and taxonomies consistently throughout this study for annotation, extraction, analysis and choice modeling. All decision rules and their corresponding illustrative examples are listed in~\ref{appendix:definition}.

\section{Data}
\label{sec:data}
\subsection{Full-text paper dataset in Transportation Research Parts}

We curated a dataset of research articles published between 2019 and 2024 in Elsevier’s Transportation Research journals, specifically Parts A through F (referred to as TR-A, TR-B to TR-F throughout the paper) and Transportation Research Interdisciplinary Perspectives (TR-IP).

\begin{figure}[htbp]
    \centering
    \includegraphics[width=\linewidth]{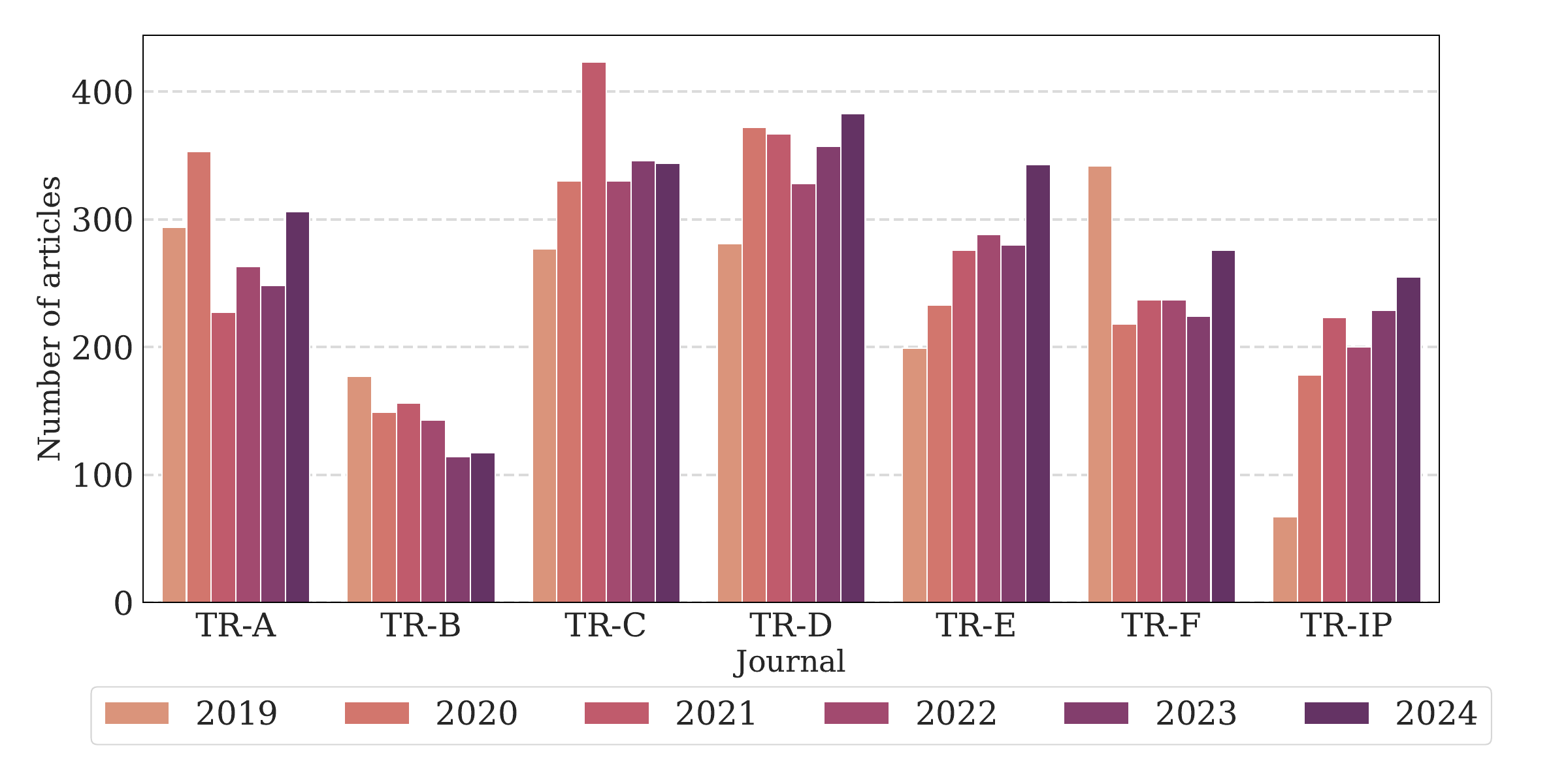}
    \caption{Number of articles published in Transportation Research journals (TR-A, TR-B, TR-C, TR-D, TR-E, TR-F, and TR-IP) from 2019 to 2024 ($N=10,724$).}
    \label{fig:overview}
\end{figure}

In total, 10,990 publications were initially extracted. The dataset consists of full-text articles in \texttt{XML} format, automatically retrieved using the Elsevier API\footnote{Text and data mining via Elsevier API, \url{https://www.elsevier.com/about/policies-and-standards/text-and-data-mining}}, which provides structured access to the content. For this study, we included only full-length research articles, excluding editorials, reviews and other non-research content. After filtering, the final dataset contained 10,724 papers. The distribution of articles by journal and year is illustrated in \figurename~\ref{fig:overview}.

The full-text dataset curation pipeline builds on Elsevier’s structured XML and stable Text-and-Data-Mining API, which makes full-text retrieval at scale straightforward and consistent across the Transportation Research journals. We focused on the TR Parts because they are flagship venues in the field. The pipeline can be transferred to other domain-specific transportation journals published by Elsevier such as Accident Analysis and Prevention, Travel Behaviour and Society, and the Journal of Transport Geography without major changes. Other transportation publishers (e.g., IEEE, Springer Nature, Wiley) provide text-mining APIs or bulk-download services~\cite{riehl2025revisiting} that allow the pipeline to be easily adapted.

\subsection{Manual annotation process}

A manual validation dataset (MVD) is used to evaluate the automated extraction of the data- and code-availability features defined in \sectionname~\ref{sec:definition}. The MVD was constructed from the manual annotation process, using papers sampled from the full-text dataset. The resulting MVD comprises 96 randomly selected full-length research articles from the set of 10,724 papers. Each paper in the MVD was independently reviewed by two annotators, who evaluated the availability of associated data and code.
We created a set of definitions and examples for each feature that are included in \ref{appendix:definition}.
The authors of this paper and other transportation researchers manually annotated the papers. Annotators received the same instructions and definitions for all annotation tasks.

\section{Automatic feature extraction}
\label{sec:feature-extraction}
\begin{figure}[htbp]
    \centering
    \includegraphics[width=0.75\linewidth]{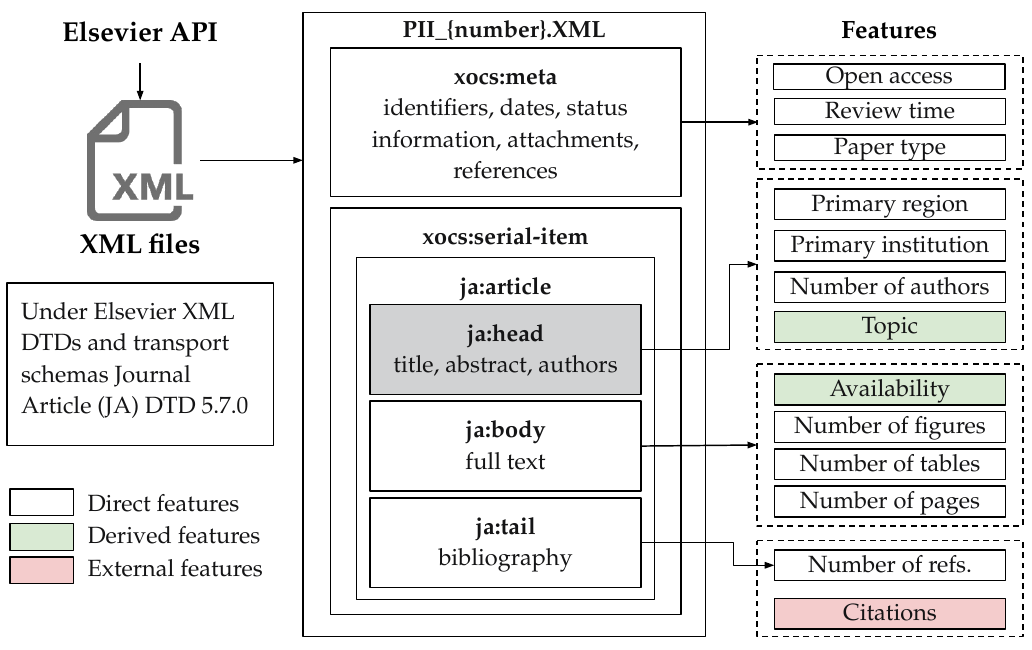}
    \caption{Overview of the features extracted from our pipeline}
    \label{fig:feature-overview}
\end{figure}

As illustrated in \figurename~\ref{fig:feature-overview}, our pipeline processes Elsevier \texttt{XML} files to generate feature sets. We parsed structural tags (e.g., \texttt{xocs:meta}, \texttt{ja:head}) to obtain direct features, while employing statistical methods and LLMs to extract derived features. Additionally, we queried the \href{https://api.elsevier.com/content/search/scopus}{Scopus Search API} to obtain citation counts, which serve as our external features.

\subsection{Article meta-data}\label{subsec:metadata_extraction}
The dataset obtained from the Elsevier API contains the full text of the papers in \texttt{XML} format. The \texttt{XML} files contain metadata about each paper (see the details of the codebook in~\cite{Elsevier_JournalsDataXML_2024}. The metadata includes information such as the title, authors, abstract, keywords, submission history, funding agencies, and more. The meta-data appears as direct features in \figurename~\ref{fig:feature-overview}. From this information, we computed the review time from initial submission to paper acceptance in days.

\subsection{Paper topic}
To categorize papers into topics, the abstracts, keywords, titles, and journal names of all full articles were processed with Latent Dirichlet Allocation (LDA), which associates words with topics~\cite{zbMATH02142783}. Given a corpus of all documents, LDA identifies topics by learning which words tend to co-occur. It represents each document as a mixture of topics and returns a probability distribution over topics for each document. Then, the highest probability topic is assigned to the document. LDA has been used previously to study the transportation research trends \cite{sun2017discovering,sun2020identifying}. More details can be found in \ref{appendix:topic}. An interactive t-SNE~\cite{maaten2008visualizing} visualization of the topic modeling results is available at \url{https://www.rerite.org/MOST/explorer}. This tool allows researchers to see the status of availability in different categories and find papers that make code or data repositories available in each category.

\begin{table}[htbp]
\caption{Topics identified by LDA }
\label{tab:topic_counts_ranked}
\centering
\footnotesize
\begin{tabular}{p{8cm}lc}
\toprule
Topic Name & Top 5 Words & Count \\
\midrule
Optimization \& Routing Algorithms & problem, model, propose, time, algorithm & 1629 \\
Social \& Policy Aspects of Mobility & study, use, mobility, policy, social & 1064 \\
Automated Driving \& Human Factors& drive, driver, vehicle, pedestrian, automate & 1004 \\
Travel \& Mode Choice Behavior & travel, choice, mode, car, trip & 922 \\
Data-Driven Modeling \& Prediction  & model, datum, base, method, use & 913 \\
Transportation Emissions \& Policy & emission, transport, fuel, port, carbon & 876 \\
Supply Chain \& Market Strategies & supply, chain, market, platform, price & 788 \\
Traffic Flow \& Network Control & traffic, vehicle, control, flow, model & 756 \\
Driver Behavior \& Safety Risk & driver, drive, behaviour, risk, safety & 651 \\
Urban Environment \& Active Transport & urban, city, parking, environment, area & 608 \\
Public Transit Service \& Demand & service, transit, passenger, bus, network & 525 \\
COVID-19 Impact on Travel \& Activities & time, covid, travel, pandemic, activity & 456 \\
Electric Vehicles \& Ride-Sharing & vehicle, electric, charge, ride, fleet & 402 \\
Air \& Freight Logistics & delivery, truck, freight, airport, air & 79 \\
Road Infrastructure \& Emergency Management & speed, road, evacuation, limit, zone & 51 \\
\bottomrule
\end{tabular}
\end{table}

After grouping all papers into 15 topics, the top words identified from each topic were given to Google Gemini 2.5 Flash to create a list of preliminary topic names. These preliminary names were then carefully reviewed and refined by the authors. The number of papers in each topic are shown in~\tablename~\ref{tab:topic_counts_ranked}.

\subsection{Features extracted by LLMs}\label{subsec:llm-extraction}
LLMs were used to extract features that are not explicitly provided in the paper meta-data. Although we initially considered text search methods, we found that text search performed worse, consistently overestimating features. Comparison of results found by LLMs and by text search can be found in \ref{appendix:just}.

We use Google Gemini model \texttt{gemini-2.5-flash-lite}~\cite{Gemini2025} via its API for the feature extraction task. It was chosen for its large maximum context size (over 1 million tokens) and because it is one of the most capable LLMs as of 2025 at the time of the feature extraction~\cite{cai2025sciassess}.

To extract the features with LLMs, the contents of the \texttt{XML} files were extracted, processed into a markdown file, and uploaded using the Google Gemini API. Multiple prompts were then given to the API to extract features about the code and data in the paper. 
Recent studies, including \cite{son2024multi,gozzi2024comparative}, demonstrate that when state-of-the-art LLMs are tasked with multiple heterogeneous subtasks in a single prompt, their accuracy declines compared to processing each instruction separately. As a result, we only extract a few related features in each prompt. Results were stored in a \texttt{JSON} file for each prompt, which are then combined. The prompt design is essential to the stability and scalability of the measurement framework. The prompts were designed to follow the operational definitions in Section~\ref{sec:definition}, to separate heterogeneous decisions into smaller and more consistent subtasks, and to return structured \texttt{JSON} outputs that could be automatically postprocessed. This design reduces ambiguity in the extraction task, makes the pipeline easier to audit, and supports repeated application to large corpora. %
Prompts can be found in~\ref{appendix:prompt} as well as repository \url{https://anonymous.4open.science/r/most-pipeline/}.
The accuracy of extracting the code and data features by LLMs is compared with human annotators in the next section.

\subsection{Human-LLM agreement evaluation}
\label{subsec:human-llm} 

\subsubsection{Evaluation criteria} 
To analyze the validity of the LLM pipeline, a three-rater inter-rater agreement (IRA) analysis~\cite{mchugh2012interrater} was conducted on the 96 papers of the MVD. The raters included two independent human annotators (H1, H2) and the automated LLM pipeline. Selecting appropriate agreement metrics requires careful consideration of our dataset characteristics. Two complementary metrics are provided. 

Our primary metric for overall agreement is Fleiss's Kappa \cite{fleiss1971}. This statistic provides a robust measure of agreement among three or more raters that corrects for the probability that agreement occurred by chance. While Fleiss's Kappa scores tend to be more conservative in datasets with high class imbalance, they offer a more meaningful assessment of true consensus. We interpret these scores using the Landis \& Koch \cite{landis1977} scale: 0.21-0.40 (\colorbox{red!15}{Fair}), 0.41-0.60 (\colorbox{orange!25}{Moderate}), 0.61-0.80 (\colorbox{yellow!40}{Substantial}), and 0.81-1.00 (\colorbox{green!20}{Almost Perfect}).

We also used Percentage Agreement (PA), which represents the raw proportion of matching annotations across raters. While intuitive and easy to interpret, PA has an important limitation in the presence of class imbalance: when negative cases heavily outnumber positive cases (as shown by the ``Prevalence'' in \tablename~\ref{tab:prev}), raters can achieve high PA simply by agreeing on the predominant negative class, without demonstrating true consensus on the more challenging positive cases. To address this, we also calculate Cohen's kappa, which measures agreement between two raters \cite{cohen1960coefficient}. Cohen's kappa ranges from -1 to 1. Landis \& Koch \cite{landis1977} suggested the same interpretation for Cohen's kappa.

\subsubsection{Agreement analysis results}
\label{sec:agreement-analysis-results}
Agreement analysis across six features revealed Fleiss' Kappa values ranging from 0.399 to 0.839, with the LLM achieving agreement levels consistently comparable to inter-human agreement. Agreement varied systematically by feature objectivity (as shown in \tablename~\ref{tab:kappa_values}). 

The most objective feature, \texttt{is\_code\_publicly\_available}, achieved \textit{Almost Perfect} agreement (Fleiss' $\kappa$ = 0.839, PA = 96.9\%), with the LLM actually exceeding human-human agreement in pairwise comparisons (Cohen's $\kappa$ = 0.928 for H1-LLM vs.\ 0.753 for H1-H2). This high agreement reflects the relatively unambiguous nature of code availability assessment. Data-related features showed more moderate results. The feature \texttt{is\_data\_cited} achieved \textit{Moderate} agreement (Fleiss' $\kappa$ = 0.522, PA = 69.8\%), while \texttt{is\_data\_repository\_available} ($\kappa$ = 0.399, PA = 91.7\%) reached \textit{Fair} agreement. Notably, the \textit{Fair} agreement score is comparable to the human-human baseline for these features (Cohen's $\kappa$ = 0.333 for data repository). This reflects genuine ambiguity in the definitions and human assessments rather than LLM-specific limitations.

For \texttt{is\_quantitative\_study}, the LLM achieved \textit{Moderate} agreement (Fleiss' $\kappa$ = 0.459), again performing comparably to human-human agreement. The high prevalence values for certain features (e.g., 96.88\% for \texttt{is\_quantitative\_study}) underscore why Kappa, rather than PA alone, provides a more meaningful assessment of consensus. 

\subsubsection{Implications} 
The agreement analysis demonstrates that our automated LLM pipeline is a valid tool for annotating artifact availability in transportation studies. The \textit{Fair} to \textit{Almost Perfect} Fleiss' Kappa scores indicate strong agreement between the LLM and two independent human experts, particularly when considering the high class imbalance in the dataset. Critically, pairwise comparisons (\tablename~\ref{tab:kappa_values}) show that the LLM's agreement with each human expert consistently matched the baseline human-human agreement, even on subjective features where the human annotators themselves showed imperfect concordance. This analysis provides confidence in using the LLM pipeline for large-scale artifact availability annotation.

\begin{table}[ht]
\caption{Percentage agreement and prevalence of true values by rater. 
Prev. = Prevalence; H1, H2 = human expert labelers 1, 2.}
\label{tab:prev}
\centering
\footnotesize
\begin{tabular}{rccccccc}
\toprule
\textbf{Features}                   & \textbf{All} & \textbf{H1 vs H2} & \textbf{H1 vs LLM} & \textbf{H2 vs LLM} & \textbf{\makecell{Prev.\\(H1)}} & \textbf{\makecell{Prev.\\(H2)}} & \textbf{\makecell{Prev.\\(LLM)}} \\ 
\midrule
\texttt{is\_code\_publicly\_available}   & 0.9688
& 0.9688
& 0.9896
& 0.9792
& 0.0833
& 0.0521
& 0.0729\\
\texttt{is\_data\_cited}
& 0.6979
& 0.8542
& 0.7500
& 0.7917
& 0.3438
& 0.3021
& 0.2604\\
\texttt{is\_data\_repository\_available} & 0.9167
& 0.9271
& 0.9271
& 0.9792
& 0.0833
& 0.0312
& 0.0312\\ 
\texttt{is\_quantitative\_study}   & 0.8854
& 0.9062
& 0.9167
& 0.9479
& 0.8854
& 0.9167
& 0.9688\\
\bottomrule
\end{tabular}
\end{table}

\begin{table}[ht]
\centering
\footnotesize
\caption{Kappa values for agreement on 96 papers. Cohen's $\kappa$ is used to measure 
agreement between 2 raters, while Fleiss's $\kappa$ can measure between multiple raters. 
Fleiss's $\kappa$ indicates how much better agreement is between annotators than would 
be expected by chance. Interpretation guide with Fleiss's $\kappa$: 0.21-0.40 (\colorbox{red!15}{Fair}), 0.41-0.60 (\colorbox{orange!25}{Moderate}), 0.61-0.80 (\colorbox{yellow!40}{Substantial}), and 0.81-1.00 (\colorbox{green!20}{Almost Perfect})~\cite{landis1977}. The background highlighting is there to make the results easier to read and interpret.} 
\label{tab:kappa_values}
\begin{tabular}{rcccc}
\toprule
\textbf{Features} & \textbf{\makecell{Cohen's $\kappa$ \\ (H1 vs H2)}} & \textbf{\makecell{Cohen's $\kappa$ \\ (H1 vs LLM)}} & \textbf{\makecell{Cohen's $\kappa$ \\ (H2 vs LLM)}} & \textbf{\makecell{Fleiss's $\kappa$ \\ (All)}} \\
\midrule
\texttt{is\_code\_publicly\_available}   & 0.7534   & 0.9277                           & 0.8226                           & \colorbox{green!20}{0.8388}                      \\
\texttt{is\_data\_cited}                & 0.6672                          & 0.4119                           & 0.4858                           & \colorbox{orange!25}{0.5224}                      \\
\texttt{is\_data\_repository\_available} & 0.3333                          & 0.3333                           & 0.6559                           & \colorbox{red!15}{0.3994}                      \\ 
\texttt{is\_quantitative\_study}        & 0.4757                          & 0.3991                           & 0.5238                           & \colorbox{orange!25}{0.4586}                      \\ 
\bottomrule
\end{tabular}
\end{table}

\subsubsection{Robustness analysis}
Because LLMs are stochastic and our temperature was not set to zero, it is possible that repeating this pipeline run would lead to different results. To verify that this performance was not an outlier, the same pipeline was run ten times on the MVD. The results are shown in Table \ref{tab:robustness-metrics}. All the Fleiss's Kappa mean values were within the same performance interval. The STDs are all under 5\% for percentage agreement and 0.05 for Fleiss's Kappa. Given the sensitivity of Fleiss's Kappa to small deviations for heavily skewed variables, this indicates that the model is performing nearly identically across runs. This is likely due to the definitions for each variable used in the prompt that reduces ambiguity even when temperature is not set to zero.

\begin{table}[ht]
\centering
\footnotesize
\caption{Inter-rater agreement metrics across dataset variables (mean $\pm$ STD)}
\label{tab:robustness-metrics}
\begin{tabular}{lcc}
\toprule
Variable & Percentage Agreement & Fleiss's Kappa \\
\midrule
\textbf{\texttt{is\_code\_publicly\_available}} & 0.9677 $\pm$ 0.0033 & 0.8316 $\pm$ 0.0202 \\
\textbf{\texttt{is\_data\_cited}} & 0.6865 $\pm$ 0.0166 & 0.4916 $\pm$ 0.0308 \\
\textbf{\texttt{is\_data\_repository\_available}} & 0.9156 $\pm$ 0.0033 & 0.3606 $\pm$ 0.0365 \\
\textbf{\texttt{is\_quantitative\_study}} & 0.8812 $\pm$ 0.0054 & 0.4635 $\pm$ 0.0196 \\
\bottomrule
\end{tabular}
\end{table}

To explore the performance of our pipeline with different models and prompting structures \cite{zhong2023chatgptunderstandtoocomparative}, we also tested them using the same pipeline and found that none of them performed uniformly better or worse than our current settings. The results can be found in \ref{appendix:diff-model}.

\subsection{Data postprocessing}
\label{subsec:postprocessing}
To support reproducible downstream analyses, we postprocess all extracted outputs into an analysis-ready dataset. The postprocessing pipeline integrates three sources: (i) structured metadata parsed from the Elsevier \texttt{XML} files (\sectionname~\ref{subsec:metadata_extraction}); (ii) LLM-extracted features returned as \texttt{JSON} files (\sectionname~\ref{subsec:llm-extraction}); and (iii) automated URL parsing and link-validation outputs derived from outbound links found in the full text and/or returned by the LLM prompts. Human annotations from the MVD (\sectionname~\ref{sec:data}) are used exclusively to evaluate extraction accuracy and agreement (\sectionname~\ref{subsec:human-llm}). %
Step-by-step postprocessing details can be found in \ref{appendix:postprocess}.

Feature inclusion for downstream analysis follows the agreement evaluation in \sectionname~\ref{subsec:human-llm}. Features with low agreement that reflect ambiguity are excluded from modeling and replaced by higher-agreement proxies as described there.
Features that remain central to the study but exhibit only Fair agreement under class imbalance (e.g., \texttt{is\_data\_repository\_available}) are retained with caution and interpreted in light of this measurement uncertainty, as discussed in \sectionname~\ref{sec:discussion}. Additional transformations required for estimation (e.g., one-hot encoding of categorical variables, collapsing sparse categories, and centering numeric covariates) are performed at the analysis stage and documented alongside the corresponding models in \sectionname~\ref{sec:analysis}. The complete list of post-processed variables, their types, and operational definitions is provided in the data dictionary (\ref{appendix:data_dict}) and implementation details are documented in the accompanying repository.

\section{Analysis}
\label{sec:analysis}

\subsection{Descriptive statistics and bivariate tests}
\begin{figure}[H]
    \centering

    \begin{minipage}{0.48\linewidth}
        \centering
        \includegraphics[width=\linewidth]{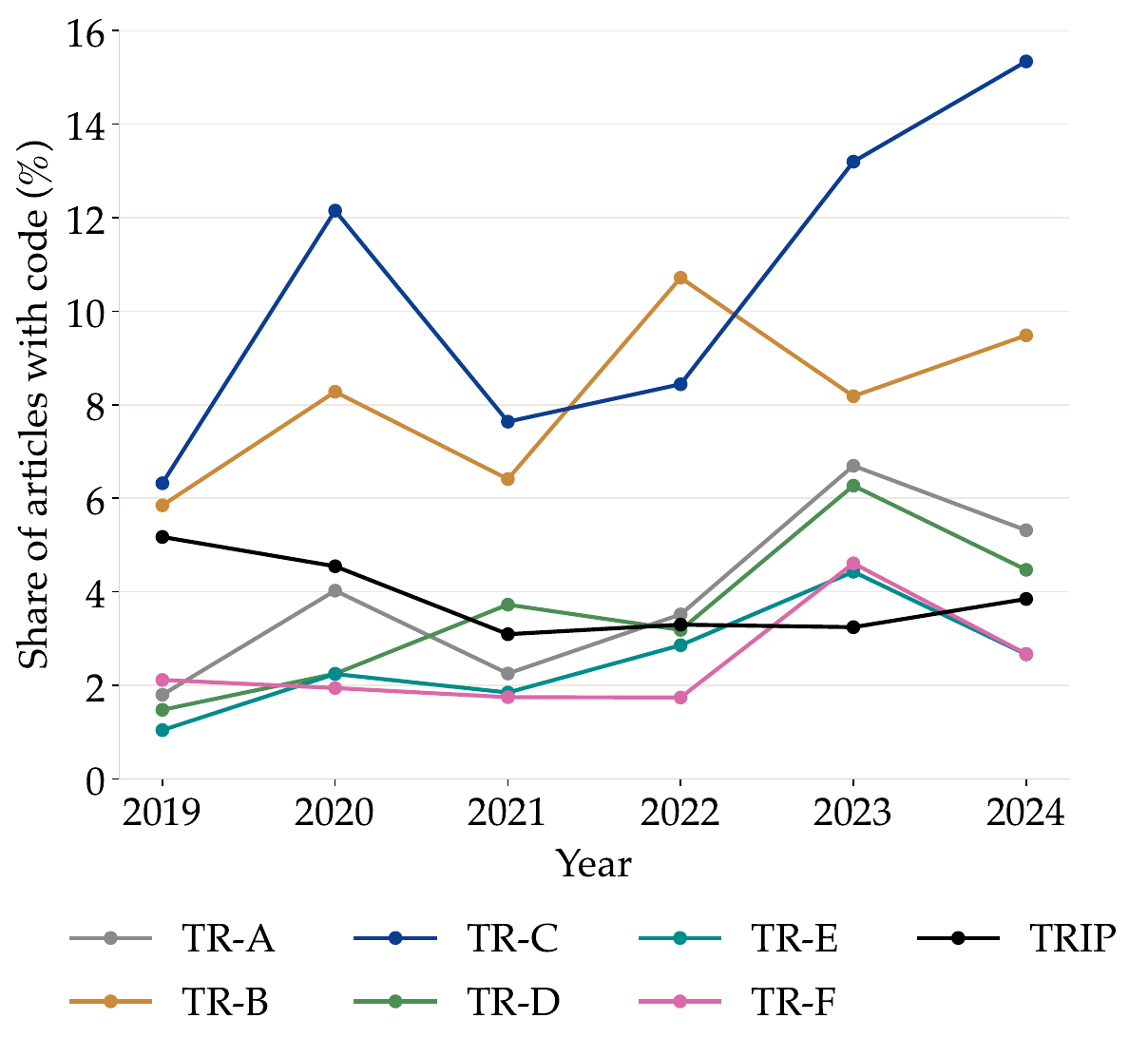}
        \caption*{(a) Code repository by journal year}
    \end{minipage}
    \hfill
    \begin{minipage}{0.48\linewidth}
        \centering
        \includegraphics[width=\linewidth]{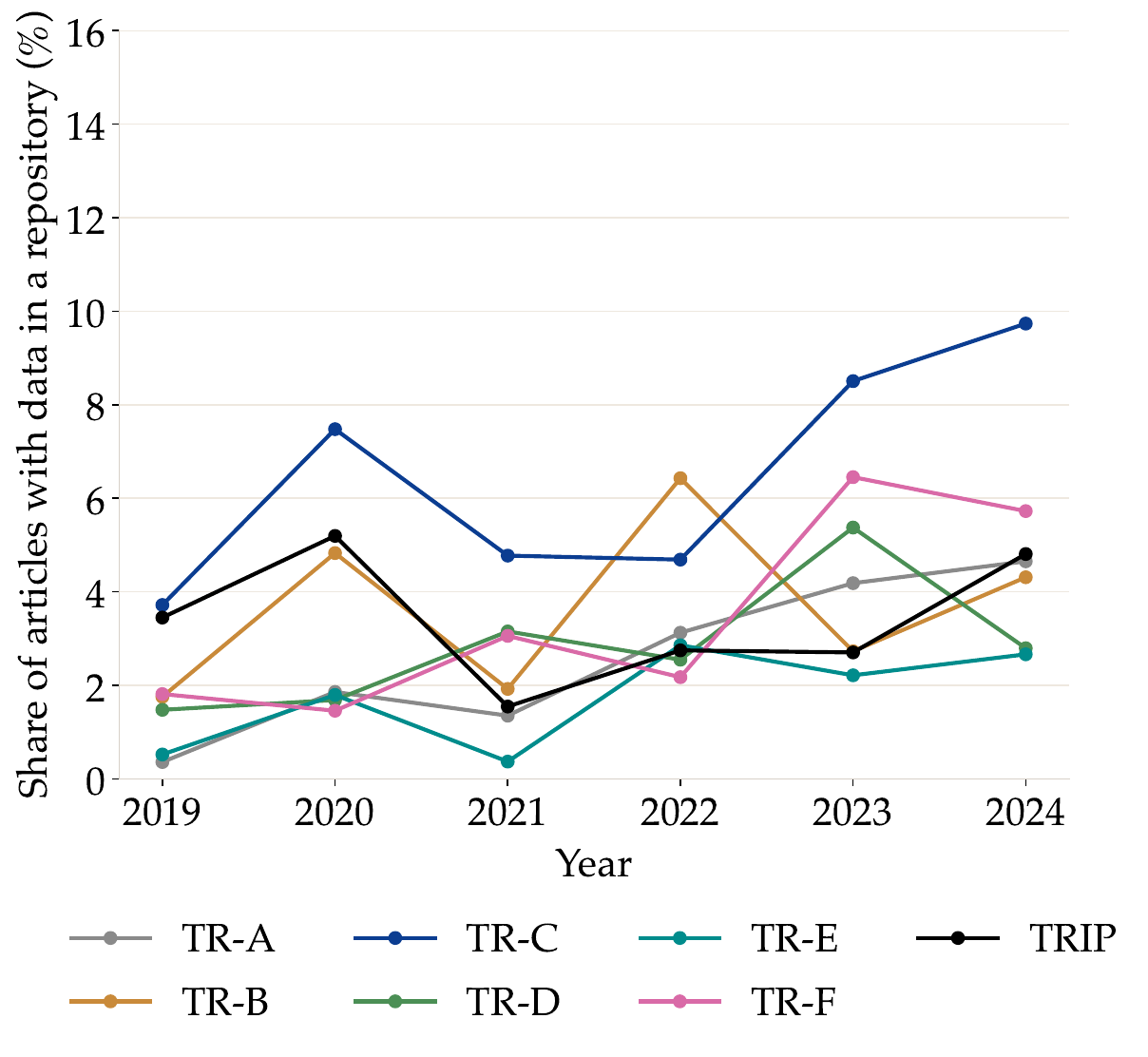}
        \caption*{(b) Data repository by journal year}
    \end{minipage}

    \caption{Comparison of data repository patterns across journals and years.}
    \label{fig:data_code_side_by_side}
\end{figure}

The analysis dataset contains the 10,480 papers that report quantitative studies (\texttt{is\_quantitative\_study}) among the full-length articles. \tablename~\ref{tab:descriptives-code}
provides descriptive statistics for these papers, organized by code availability. The dataset contains 528 papers with code available, a little more than 5\% of the dataset. 
In total, about
29\% of the papers cite available data, but only 4\% of the
papers share a data repository; the remaining 67\% neither cite data
nor include a repository. \figurename~\ref{fig:data_code_side_by_side} presents the temporal trends in data and code sharing within our
research community over the years covered by this study.

\begin{table}[!htbp]
\centering
\caption{Descriptive statistics of variables by code availability} %
\label{tab:descriptives-code}
\footnotesize
\begin{tabular}[t]{llrrrr}
\toprule
\multicolumn{2}{c}{ } & \multicolumn{2}{c}{Code available ($N=528$)} & \multicolumn{2}{c}{Not available ($N=9952$)} \\
\cmidrule(l{3pt}r{3pt}){3-4} \cmidrule(l{3pt}r{3pt}){5-6}
\multicolumn{2}{l}{\textbf{Numeric variables}} & \multicolumn{1}{c}{Mean} & \multicolumn{1}{c}{Std. Dev.} & \multicolumn{1}{c}{Mean} & \multicolumn{1}{c}{Std. Dev.}\\
\cmidrule(l{3pt}r{3pt}){1-2} \cmidrule(l{3pt}r{3pt}){3-4} \cmidrule(l{3pt}r{3pt}){5-6}
Age of paper [years] &  & 3.1 & 1.7 & 3.4 & 1.7\\
N. tables in paper &  & 6.3 & 4.6 & 6.4 & 4.5\\
N. figures in paper &  & 11.0 & 6.5 & 8.6 & 6.1\\
N. references &  & 57.3 & 21.0 & 58.2 & 26.9\\
N. authors &  & 3.8 & 1.7 & 3.8 & 1.7\\ 
Times cited &  & 27.2 & 42.9 & 28.4 & 43.2\\
Review time [days] &  & 268.6 & 141.4 & 254.3 & 148.9\\
Page count &  & 19.8 & 6.6 & 18.4 & 6.3\\
\midrule
\multicolumn{2}{l}{\textbf{Categorical variables}} & \multicolumn{1}{c}{N} & \multicolumn{1}{r}{\%} & \multicolumn{1}{c}{N} & \multicolumn{1}{r}{\%}\\
\midrule
Data availability & Data is cited or linked & 96 & 18.2 & 2810 & 28.2\\
 & Data repository and citation & 94 & 17.8 & 14 & 0.1\\
 & Data repository is available & 199 & 37.7 & 65 & 0.7\\
 & No repository or citation & 139 & 26.3 & 7063 & 71.0\\
\midrule
Open Access & Not open access & 324 & 61.4 & 7346 & 73.8\\
 & Open access & 204 & 38.6 & 2606 & 26.2\\
\midrule
Journal & TR-IP & 36 & 6.8 & 945 & 9.5\\
 & TR-A & 64 & 12.1 & 1555 & 15.6\\
 & TR-B & 67 & 12.7 & 771 & 7.7\\
 & TR-C & 212 & 40.2 & 1797 & 18.1\\
 & TR-D & 72 & 13.6 & 1911 & 19.2\\
 & TR-E & 41 & 7.8 & 1534 & 15.4\\
 & TR-F & 36 & 6.8 & 1439 & 14.5\\
\midrule
Region of corr. author & Africa & 5 & 0.9 & 69 & 0.7\\
 & Asia & 118 & 22.3 & 4398 & 44.2\\
 & Europe & 206 & 39.0 & 2699 & 27.1\\
 & North America & 156 & 29.5 & 2074 & 20.8\\
 & Oceania & 27 & 5.1 & 521 & 5.2\\
 & South America & 16 & 3.0 & 191 & 1.9\\
\midrule
Availability statement & Availability statement is not present & 139 & 26.3 & 7062 & 71.0\\
 & Availability statement is present & 389 & 73.7 & 2890 & 29.0\\
\midrule
Paper topic & Air \& Freight Logistics & 2 & 0.4 & 77 & 0.8\\
 & Automated Driving \& Human Factors & 34 & 6.4 & 957 & 9.6\\
 & COVID-19 Impact on Travel \& Activities& 17 & 3.2 & 434 & 4.4\\
 & Data-Driven Modeling \& Prediction & 142 & 26.9 & 772 & 7.8\\
 & Driver Behavior \& Safety Risk & 10 & 1.9 & 631 & 6.3\\
 & Electric Vehicles \& Ride-Sharing & 29 & 5.5 & 372 & 3.7\\
 & Optimization \& Routing Algorithms & 114 & 21.6 & 1513 & 15.2\\
 & Public Transit Service \& Demand & 29 & 5.5 & 489 & 4.9\\
 & Road Infrastructure \& Emergency Management & 1 & 0.2 & 49 & 0.5\\
 & Social \& Policy Aspects of Mobility  & 26 & 4.9 & 862 & 8.7\\
 & Supply Chain \& Market Strategies  & 6 & 1.1 & 775 & 7.8\\
 & Traffic flow \& Network Control & 45 & 8.5 & 710 & 7.1\\
 &Transportation Emissions \& Policy & 20 & 3.8 & 848 & 8.5\\
 & Travel \& Mode Choice Behavior & 28 & 5.3 & 883 & 8.9\\
 & Urban Environment \& Active Transport  & 25 & 4.7 & 580 & 5.8\\
\midrule
Links to code repositories & Link to GitHub & 329 & 62.3 & 4 & 0.0\\
& Link to other service & 118 & 22.3 & 5 & 0.1\\
& No links & 81 & 15.3 & 9943 & 99.9\\
\bottomrule
\end{tabular}
\end{table}

We conducted a series of bivariate statistical tests on the dependent variables, specifically, whether the papers provide code or data, using $t$-tests on differences of means when the independent variable is numeric, and Pearson $\chi^2$ tests of independence when the independent variable is categorical. The results of these tests revealed several statistically significant relationships that guided the analysis that follows in this section. To begin, papers that make code available are also more likely to share data. Newer papers make code available more frequently, and newer papers also tend to share data directly rather than simply cite existing
data. Papers with corresponding authors in Europe and North America are more likely to make data available than papers with authors in Asia; the same trend holds for sharing code, with South American authors joining the group more likely to share than Asian authors. In absolute terms, authors based in Asia still contribute a substantial number ($N=118$) of code repositories, but their share appears lower when expressed as a percentage of the total number of publications from the region ($N=4516$). 
The usage of data varies widely by topic, with the topics "Urban Environment \& Active Transport" and "COVID-19 Impact on Travel \& Activities" 
having higher than expected rates of citing data from others (about 45\% of papers cited or linked to external datasets in these topic categories). Papers in the topic "Data-driven modeling and prediction" share code at a much higher rate than other topics (about 15\% of papers have code available). Other comparisons did not result in bivariate test statistics sufficient to reject a null hypothesis of no relationship or no difference in means. Note that the numbers reported in the analysis section are subject to estimation error arising from the methodology.

These findings are informative, but they may be the result of correlated omitted variables. A complete list of statistics of all variables can be viewed \href{https://019b33f2-34f4-9d17-cc65-86ffbaaa9028.share.connect.posit.cloud/}{here}.
In the next section we provide a choice model of data and code availability that
can accommodate the simultaneous influence of multiple variables.

\subsection{Choice model predicting the availability of code}

The factors predicting the availability of code were analyzed using a logit model. This model allows one to investigate the effect of several explanatory variables while controlling for possible confounding factors. The utility functions for code available (CA) and code unavailable (CU) for each paper $n$ are presented in equations~\eqref{eq:CA} and~\eqref{eq:CU}:

\begin{align}
U^{CA}_{n} &= \alpha^{CA} + \boldsymbol{\beta}^{CA} \mathbf{X}^{CA}_{n} + \varepsilon^{CA}_{n}, \label{eq:CA} \\
U^{CU}_{n} &= 0 + \varepsilon^{CU}_{n}, \label{eq:CU}
\end{align}
where $\alpha^{CA}$ is the constant,  $\boldsymbol{\beta}^{CA}$ is the vector of parameters associated with the explanatory variables $\mathbf{X}^{CA}_{n}$, and $\varepsilon^{CA}_{n}$ and $\varepsilon^{CU}_{n}(t)$ are i.i.d. extreme value distributed error terms. Given the choice indicator $Y_{n}$, the probability of code being available is given by equation~\eqref{eq:probCA}:

\begin{equation}
P_{n}\!\big(Y_{n} = 1\big)
= \frac{
\exp\!\left( \alpha^{CA} + \boldsymbol{\beta}^{CA}\mathbf{X}^{CA}_{n} \right)
}{
1 + \exp\!\left( \alpha^{CA} + \boldsymbol{\beta}^{CA}\mathbf{X}^{CA}_{n} \right)
}.
\label{eq:probCA}
\end{equation}\textbf{}

The parameters were estimated using the Biogeme Python package ~\cite{bierlaire2023biogeme}. The explanatory variables presented in the final model specification were chosen according to interpretation and statistical significance (p-value $<$ 0.05). The numerical explanatory variables were centered on the means. Categorical variables were coded as multiple binary variables. The most frequent category was assumed to be the reference category and fixed to zero.  When a variable was available for part of the papers, a binary variable was included into the equation to denote the missing values.  The effect of each explanatory variable was tested using the likelihood ratio test between different model specifications. All variables related to the journal, the region of the corresponding author, the paper topic, and the paper characteristics were initially included in the utility function. The variables that showed low significance according to descriptive statistics were tested first. Binary variables extracted from the same categorical variable that had a similar impact on code availability were combined into a new variable. Variables that did not have a significant effect were excluded one by one.

\tablename~\ref{tab:model_statistics_code} shows the goodness of fit measures for the final model and for a baseline model that includes only a constant. The adjusted likelihood ratio index shows that the inclusion of explanatory variables resulted in an improvement of 11.2\% compared to the model that includes only a constant.

\begin{table}[htbp]
\centering
\caption{Statistics for the choice model predicting the availability of code}
\small
\begin{tabular}{lr}
\toprule
\textbf{Statistic} & \textbf{Value} \\
\midrule
Number of parameters associated with explanatory variables ($K$) & 8 \\
Number of observations & 10,480 \\
Constant log likelihood ($L(c)$) & -2,092 \\
Final log likelihood  ($L(\hat{\beta}) $) & -1,850 \\
Adjusted likelihood ratio index $\rho^2 = 1 - \frac{L(\hat{\beta}) - K}{L(c)}$ & 0.112 \\
\bottomrule
\end{tabular}
\label{tab:model_statistics_code}
\end{table}

\tablename~\ref{tab:results_code} shows the estimation results. All parameters are statistically significant at the 5\% level. The alternative specific constant is negative, meaning that the papers are less likely to make code available, everything else being equal.

\begin{table}[htbp]
\centering
\caption{Choice model predicting the availability of code}
\begin{tabular}{p{3.5cm}p{8.5cm}c}
\toprule
\textbf{Variable} & \textbf{Description} & \textbf{Estimate (Robust SE)} \\
\midrule

$\alpha$ & Alternative-specific constant 
& -3.89 (0.124) *** \\
TRB, TRC & Binary variable equal to one when the journal is TR-B or TR-C 
& 0.785 (0.104) *** \\
Europe, SouthAmerica, Africa & Binary variable equal to one when the region is Europe, South America, or Africa 
& 1.37 (0.122) *** \\
NorthAmerica, Oceania & Binary variable equal to one when the region is North America or Oceania 
& 1.11 (0.125) *** \\
DataDriven & Binary variable equal to one when the topic is data-driven methods 
& 0.738 (0.123) *** \\
AutoDri, Cov19, SocPol, TraFlow, TraEm, TraMo & Binary variable equal to one when the topic is automated driving, COVID-19, transport policy, traffic flow, transport emissions, or travel mode 
& -0.613 (0.113) *** \\
DrivBeh, Log, RoInf, SuCha & Binary variable equal to one when the topic is driver behavior, logistics, road infrastructure, or supply chain 
& -1.32 (0.248) *** \\
NumFig & Number of figures in the paper 
& 0.0194 (0.00738) ** \\
PaperAge & Paper age in years 
& -0.178 (0.0284) *** \\
\bottomrule
\end{tabular}

\begin{tablenotes}
\small
\item \textit{Notes:} Robust standard errors in parentheses. 
*** $p < 0.001$, ** $p < 0.01$, * $p < 0.05$.
\end{tablenotes}

\label{tab:results_code}
\end{table}

Certain journals had a significant effect on code availability. Papers published in TR-D were initially chosen as the reference category. Papers published in TR-B and TR-C were more likely to have code available. The effect was not significantly different between the two journals. Papers in the other journals (TR-A, TR-E, TR-F, and TR-IP) did not have a significant effect compared to TR-D.

The regions in which the first author was located had a significant effect on code availability. Papers with authors in Europe, South America, Africa, North America, and Oceania were more likely to have code available than papers with authors in Asia. This effect was significantly higher for Europe, South America and Africa than for North America and Oceania. The effect was not significantly different within the group Europe, South America, and Africa and within the group North America and Oceania. 

The topics of the papers had a significant impact on code availability. Papers on optimization were initially chosen as the reference category. Papers based on data-driven methods were the most likely to have available code. Papers focusing on automated driving, COVID-19, transport policy, traffic flow, transport emissions, and travel mode were less likely to have code available. The effect was not significantly different between these topics. Topics such as driver behavior, logistics, road infrastructure, and supply chain were the least likely to have code available. The effect was not significantly different between these topics. The other topics did not have a significant effect compared to optimization. 

Certain characteristics of the paper had a significant effect on the availability of code. Papers with a high number of figures were more likely to have available code. Older papers were less likely to have code available. The number of authors, number of tables, reference count, and page count were tested but did not exhibit a significant effect and were therefore excluded from the final model specification.

\subsection{Choice model predicting the availability of data}

The factors predicting the availability of data were analyzed using a logit model. The utility functions of data citation and data repository (C, R), no data citation and data repository (NC, R), data citation and no data repository (C, NR), no data citation and no data repository (NC, NR) for each paper $n$ are presented in equations~\eqref{eq:C,R},~\eqref{eq:NC,R},~\eqref{eq:C,NR} and~\eqref{eq:NC, NR}:

\begin{align}
U^{C,R}_{n} &= \alpha^{C,R} + \boldsymbol{\beta}^{C,R} \mathbf{X}^{C,R}_{n} + \varepsilon^{C,R}_{n}, \label{eq:C,R} \\
U^{NC,R}_{n} &= \alpha^{NC,R} + \boldsymbol{\beta}^{NC,R} \mathbf{X}^{NC,R}_{n} + \varepsilon^{NC,R}_{n}, \label{eq:NC,R} \\
U^{C,NR}_{n} &= \alpha^{C,NR} + \boldsymbol{\beta}^{C,NR} \mathbf{X}^{C,NR}_{n} + \varepsilon^{C,NR}_{n}, \label{eq:C,NR} \\
U^{NC, NR}_{n} &= 0 + \varepsilon^{NC, NR}_{n}, \label{eq:NC, NR}
\end{align}
where $\alpha^{C,R}$, $\alpha^{NC,R}$, $\alpha^{C,NR}$ are the alternative specific constants,  $\boldsymbol{\beta^{C,R}}$, $\boldsymbol{\beta^{NC,R}}$, $\boldsymbol{\beta^{C,NR}}$  are the vectors of parameters associated with the explanatory variables $\mathbf{X}^{C,R}_{n}$, $\mathbf{X}^{NC,R}_{n}$, $\mathbf{X}^{C,NR}_{n}$ and $\varepsilon^{C,R}_{n}$, $\varepsilon^{NC,R}_{n}$, $\varepsilon^{C,NR}_{n}$ are i.i.d. extreme value distributed error terms. Given the choice indicator $Y_{n}$, the probability that data is available $k \in C_{D}$ with 
$C_{D} = \{\text{C,R};\, \text{NC,R};\, \text{C,NR};\, \text{NC,NR}\}$ is given by equation~\eqref{eq:probDA}:

\begin{equation}
P_{n}\!\big(Y_{n} = k\big)
= 
\frac{
\exp\!\left( \alpha^{k} + \boldsymbol{\beta}^{k}\mathbf{X}^{k}_{n} \right)
}{
\displaystyle\sum_{j \in C_{D}}
\exp\!\left( \alpha^{j} + \boldsymbol{\beta}^{j}\mathbf{X}^{j}_{n} \right)
}.
\label{eq:probDA}
\end{equation}\textbf{}

The parameters were estimated using the Biogeme Python package~\cite{bierlaire2023biogeme}. The explanatory variables presented in the final model specification were chosen according to interpretation and statistical significance (p-value $<$ 0.05). The explanatory variables were processed as described in the previous section.
All variables related to the journal, the region of the corresponding author, the paper topic, and the paper characteristics were initially included in the utility functions and assumed to have the same impact between the alternatives. Alternative specific parameters for each variable were tested using the likelihood ratio test. The variables that showed high significance based on descriptive statistics were first tested. The variables were combined and excluded as described in the previous section. Cross-nested and nested logit specifications were also tested to capture unobserved correlations between alternatives that include data citation and/or data repository. The estimation results did not produce a significant improvement in the final log likelihood. The nesting and membership parameters were not statistically significant. Therefore, the logit model was retained as the final specification. 

\tablename~\ref{tab:model_statistics_data} shows the goodness of fit measures for the final model and for a baseline model that includes only a constant. The adjusted likelihood ratio index shows that the inclusion of explanatory variables resulted in an improvement of 9.2\% compared to the model that includes only a constant. 

\begin{table}[htbp]
\centering
\caption{Statistics for the choice model predicting the availability of data}
\small
\begin{tabular}{l r}
\toprule
\textbf{Statistic} & \textbf{Value} \\
\midrule
Number of parameters associated with explanatory variables ($K$) & 26 \\
Number of observations & 10,480 \\
Constant log likelihood ($L(c)$) & -7,895 \\
Final log likelihood ($L(\hat{\beta})$) & -7,146 \\
Adjusted likelihood ratio index $\rho^2 = 1 - \frac{L(\hat{\beta}) - K}{L(c)}$ &  0.0916\\
\bottomrule
\end{tabular}
\label{tab:model_statistics_data}
\end{table}

\tablename~\ref{tab:results_data} shows the estimation results. Almost all parameters are statistically significant at the 5\% level. One parameter that was significant at the 10\% level was retained because it resulted in a significant improvement based on the likelihood ratio test. The alternative-specific constants are negative, meaning that the papers are less likely to provide a data citation or a data repository, everything else being equal. This effect is larger for providing a data repository than for providing a data citation.

Certain journals had a significant effect on data availability. Papers published in TR-D were initially chosen as the reference category. Papers in TR-B and TR-E were less likely to have a data citation and no data repository. The effect was not significantly different between the two journals. Papers in TR-C were more likely to have a data repository, regardless of data citation. Papers published in TR-F were less likely to have a data citation, regardless of the availability of a data repository. Papers published in TR-IP were less likely to have a data citation and a data repository. The other journals did not have a significant effect compared to TR-D.

The region in which the first author was located had a significant effect on data availability. Asia was initially chosen as the reference category. Papers with authors in Africa, North America, Europe, and Oceania were more likely to have a data citation and no data repository. This effect was significantly higher for Africa and North America than for Europe and Oceania. Articles with authors in Europe, North America, Oceania, and South America were more likely to provide a data repository regardless of data citation. This effect was significantly higher for Europe and North America than for Oceania and South America. The other regions did not have a significant effect compared to Asia.

\begin{table}[H]
\centering
\caption{Choice model predicting the data citation and the data repository}
\resizebox{\textwidth}{!}{
\begin{tabular}{p{3.5 cm}p{8.5 cm}ccc}
\toprule
\textbf{Variable} & \textbf{Description} & \multicolumn{3}{c}{\textbf{Estimate (Robust SE)}} \\
\cmidrule(lr){3-5}
 & & $V_{C,NR}$ & $V_{NC,R}$ & $V_{C,R}$ \\
\midrule
$\alpha$ & Alternative-specific constants & -0.892 (0.0580) *** & -4.46 (0.149) *** & -5.60 (0.193) *** \\
TRB, TRE & Binary variable equal to one when the journal is TR-B or TR-E & -0.263 (0.0640) *** & -- & -- \\
TRC & Binary variable equal to one when the journal is TR-C & -- & 0.556 (0.128) *** &  0.556 (0.128) ***\\
TRF & Binary variable equal to one when the journal is TR-F & -1.23 (0.117) *** & -- & -1.23 (0.117) *** \\
TRIP & Binary variable equal to one when the journal is TR-IP & -- & -- & -0.859 (0.431) * \\
Africa, NorthAmerica & Binary variable equal to one when the region is Africa or North America & 0.769 (0.0592) *** & -- & -- \\
Europe, Oceania & Binary variable equal to one when the region is Europe or Oceania & 0.351 (0.0560) *** & -- & -- \\
Europe, NorthAmerica & Binary variable equal to one when the region is Europe or North America & -- & 1.68 (0.142) *** & 1.68 (0.142) *** \\
Oceania, SouthAmerica & Binary variable equal to one when the region is Oceania or South America & -- & 1.17 (0.237) *** & 1.17 (0.237) *** \\
DataDriven & Binary variable equal to one when the topic is data-driven methods & 0.355 (0.0867) *** & 0.879 (0.174) *** & 1.83 (0.227) *** \\
Cov19, UrbEnv & Binary variable equal to one when the topic is COVID-19 or urban environment & 0.443 (0.0812) *** & -- & 1.16 (0.284) *** \\
TransEmiss & Binary variable equal to one when the topic is transport emissions & 0.335 (0.0862) *** & -0.525 (0.275) & -0.525 (0.275) \\
DrivBeh & Binary variable equal to one when the topic is driver behavior & -0.297 (0.125) * & -0.297 (0.125) * & -0.297 (0.125) * \\
SocPol, TraFlow & Binary variable for transport policy and traffic flow topics &  -0.628 (0.0751) *** & -0.628 (0.0751) *** &  -0.628 (0.0751) *** \\
AutoDri & Binary variable equal to one when the topic is automated driving & -1.05 (0.140) *** & -- & --  \\
TraMo, Log, RoInf, SuCha & Binary variable for travel mode, logistics, road infrastructure, or supply chain topics & -0.504 (0.0742) *** & -1.16 (0.238) *** & -1.16 (0.238) *** \\
NumFig & Number of figures in the paper & 0.0238 (0.00507) *** & 0.0238 (0.00507) *** & 0.0238 (0.00507) *** \\
NumTab & Number of tables in the paper & 0.0197 (0.00589) *** & 0.0197 (0.00589) *** & -- \\
NumPag & Number of pages & -0.0383 (0.00584) *** & -0.0383 (0.00584) *** & -0.0383 (0.00584) *** \\
NumAuth & Number of authors & -0.0340 (0.0142) * & -- & -- \\
PaperAge & Paper age in years & 0.0740 (0.0139) *** &  -0.224 (0.0337) ***  & -0.224 (0.0337) *** \\
\bottomrule
\end{tabular}
}
\begin{tablenotes}
\small
\item \textit{Notes:} Robust standard errors in parentheses. 
*** $p < 0.001$, ** $p < 0.01$, * $p < 0.05$. 
`--` indicates that the variable is not included in the corresponding utility function.
For variables with identical estimates across utilities, equality restrictions were imposed based on statistical tests.
\end{tablenotes}

\label{tab:results_data}
\end{table}

The topic of the paper had a significant impact on data availability. Papers on optimization were initially chosen as the reference category. Papers based on data-driven methods were the most likely to provide a data citation and/or a data repository. The effect was higher when both a data citation and a data repository were provided. Papers focusing on COVID-19 or urban environment were more likely to have a data citation, and the effect was higher when a repository was also provided. The effect was not significant for a data repository without a data citation. The effect was not significantly different for these two topics. Papers on transport emissions were more likely to have a data citation without a data repository and less likely to provide a repository.
Papers on driver behavior, transport policy, and traffic flow were less likely to provide a data citation and/or a data repository. The effect was higher for transport policy and traffic flow than for driver behavior. The effect did not differ significantly across a data citation and/or a data repository. 
Papers on automated driving were less likely to have a data citation and no data repository.
Topics such as travel mode, logistics, road infrastructure, and supply chain were the least likely to provide a data citation and/or a data repository. The effect was higher for papers providing a data repository. The effect was not significantly different between these topics. The other topics did not have a significant effect compared to optimization. 

Certain characteristics of the paper had a significant effect on the availability of data. Papers with a high number of figures were more likely to provide a data citation and/or a data repository. The effect did not differ significantly between a data citation and a data repository. Papers with a high number of tables were more likely to have either a data citation or a data repository, but not both. The effect did not differ significantly between a data citation and a data repository. Papers with a high number of pages were less likely to have data available. The effect did not differ significantly between a data citation and a data repository. 

Papers with a high number of authors were less likely to have a data citation and no data repository. The effect of the number of authors on the other alternatives was not significant. Older papers were more likely to have a data citation and no data repository and less likely to provide a repository. 
The reference count did not have a significant effect.

\subsection{Incentives for availability}

Open science techniques often require additional investment and effort by researchers. If the discipline does not require these efforts, researchers may respond to various incentives. In our data, two incentives that we capture include a reduced review time prior to publication or a higher rate of citations on the resulting publication. The review time for a paper can be determined in the paper metadata by calculating the number of days from when the paper was first submitted to when the paper was accepted for publication, while the rate of citations is calculated as the number of citations in SCOPUS divided by the whole number of years since publication. Unfortunately, neither incentive is observable in the papers we studied.

\begin{figure}[htpb]
  \centering

  \begin{subfigure}[t]{0.48\linewidth}
    \centering
    \includegraphics[width=\linewidth]{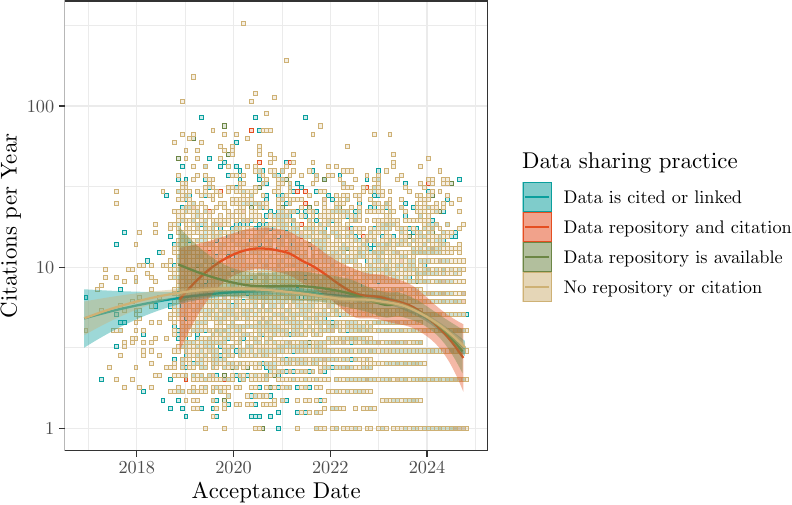}
    \caption{Citations per year by data sharing practice.}
    \label{fig:citations-data}
  \end{subfigure}
  \begin{subfigure}[t]{0.48\linewidth}
    \centering
    \includegraphics[width=\linewidth]{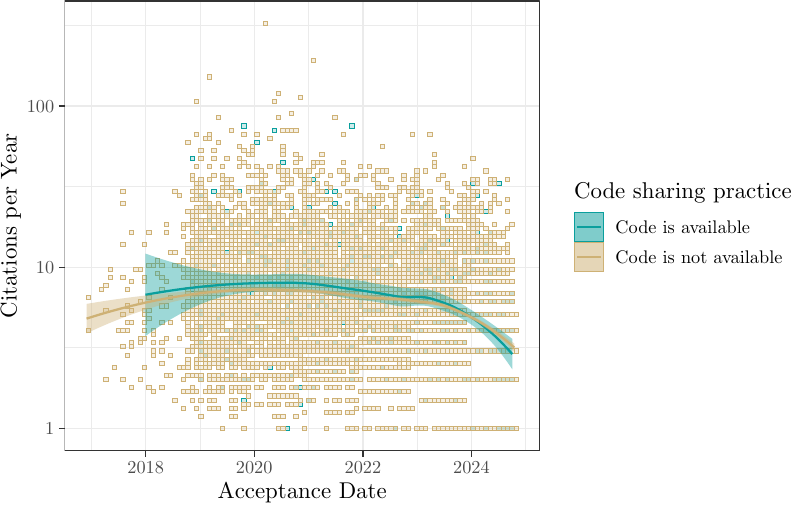}
    \caption{Citations per year by code sharing practice.}
    \label{fig:citations-code}
  \end{subfigure}

  \caption{Citations per year of each paper over time by code and data-sharing practice, with a LOESS moving average line applied for mean comparison.}
  \label{fig:citations}
\end{figure}

In terms of the number of citations, Figure \ref{fig:citations} shows the share of citations per year for each paper
ordered by paper acceptance date in SCOPUS.  Figure \ref{fig:citations-data} groups the papers by their data availability in the categories described above, and Figure
\ref{fig:citations-code} groups the papers by code availability. 
Both figures show a LOESS moving average rate of the mean citation rate per year, along with a 95\% confidence interval of the mean. In
neither case are the average citation rates different based on the availability of data or code. 
Figure \ref{fig:citations-data} shows a temporary increase in the citation rate for papers with a data repository and citation in 2020 and 2021, to the point where the uncertainty interval does not overlap the uncertainty interval of the other papers with different data sharing practices. Further examination of this effect showed that two papers with unusually high citations drive this effect. There was no such temporal variability in the citations or review time grouped by data sharing practice.%

Table~\ref{tbl-incentives} presents estimates of log-linear regression
models of the review time in days and citations per year for the papers
in the dataset. These models include the journal, paper topic, and
region of corresponding author as dummy controls to account for
differing practices between journals and reviewers in topics.
Insignificant parameter estimates on these controls are suppressed in
Table~\ref{tbl-incentives} for clarity but the controls by themselves
explain virtually all of the reduction in variance in the models, with
insignificant reductions in the model \(F\)-statistic when markers for
data sharing and code availability are included (\(p\)-value of 0.44 for
review time and 0.33 for citations per year).

\begin{table}[!htbp]
\centering
\caption{Incentives for Open Science}
\label{tbl-incentives}
\resizebox{\textwidth}{!}{
\begin{threeparttable}
\begin{tabular}{lcccc}
\toprule
 & \multicolumn{2}{c}{Review time [days]} & \multicolumn{2}{c}{Citations per year} \\
\cmidrule(lr){2-3} \cmidrule(lr){4-5}
 & Controls & Open Science & Controls & Open Science \\
\midrule
Intercept & 5.189*** & 5.198*** & 1.262*** & 1.273*** \\
& [5.010, 5.369] & [5.019, 5.378] & [1.022, 1.503] & [1.032, 1.514] \\
Code is available &  & 0.013 &  & -0.067 \\
&  & [-0.051, 0.076] &  & [-0.154, 0.020] \\
Data repository is available (NC, R) &  & -0.018 &  & 0.020 \\
&  & [-0.099, 0.064] &  & [-0.092, 0.132] \\
Data is cited or linked (C, NR) &  & -0.019 &  & -0.012 \\
&  & [-0.045, 0.007] &  & [-0.046, 0.022] \\
Data repository and citation (C, R) &  & 0.052 &  & 0.138+ \\
&  & [-0.068, 0.171] &  & [-0.023, 0.299] \\
Journal: TR - A & 0.755*** & 0.754*** & 0.415*** & 0.415*** \\
& [0.710, 0.800] & [0.709, 0.800] & [0.353, 0.476] & [0.353, 0.476] \\
Journal: TR - B & 0.619*** & 0.616*** & 0.523*** & 0.524*** \\
& [0.563, 0.675] & [0.560, 0.672] & [0.446, 0.601] & [0.446, 0.602] \\
Journal: TR - C & 0.415*** & 0.413*** & 0.722*** & 0.722*** \\
& [0.368, 0.462] & [0.366, 0.460] & [0.656, 0.787] & [0.657, 0.788] \\
Journal: TR - D & 0.033 & 0.032 & 0.600*** & 0.599*** \\
& [-0.017, 0.083] & [-0.018, 0.081] & [0.538, 0.661] & [0.538, 0.661] \\
Journal: TR - E & 0.305*** & 0.303*** & 0.682*** & 0.680*** \\
& [0.253, 0.356] & [0.252, 0.355] & [0.610, 0.753] & [0.609, 0.752] \\
Journal: TR - F & 0.223*** & 0.220*** & 0.231*** & 0.228*** \\
& [0.171, 0.274] & [0.168, 0.271] & [0.159, 0.303] & [0.156, 0.300] \\
Topic: COVID-19 Impact on Travel \& Activities & -0.213** & -0.213** & 0.020 & 0.019 \\
& [-0.347, -0.080] & [-0.347, -0.080] & [-0.160, 0.200] & [-0.162, 0.199] \\
Topic: Social \& Policy Aspects of Mobility & -0.097 & -0.100 & 0.265** & 0.262** \\
& [-0.225, 0.030] & [-0.228, 0.028] & [0.091, 0.438] & [0.088, 0.435] \\
Topic: Electric Vehicles \& Ride-Sharing & -0.035 & -0.035 & 0.206* & 0.208* \\
& [-0.170, 0.100] & [-0.170, 0.100] & [0.025, 0.388] & [0.026, 0.389] \\
Topic: Road Infrastructure \& Emergency Management & 0.197+ & 0.195+ & -0.270* & -0.271* \\
& [-0.019, 0.412] & [-0.021, 0.411] & [-0.537, -0.003] & [-0.538, -0.005] \\
Region: South America & 0.047 & 0.044 & -0.209* & -0.212* \\
& [-0.105, 0.199] & [-0.108, 0.197] & [-0.409, -0.009] & [-0.412, -0.013] \\
\midrule
Num.Obs. & 9185 & 9185 & 10480 & 10480 \\
R2 & 0.182 & 0.183 & 0.113 & 0.114 \\
R2 Adj. & 0.180 & 0.180 & 0.111 & 0.111 \\
Log.Lik. & -7138.461 & -7136.564 & -11825.901 & -11823.609 \\
RMSE & 0.53 & 0.53 & 0.75 & 0.75 \\
\bottomrule
\end{tabular}

\begin{tablenotes}
\footnotesize
\item 95\% confidence intervals are reported in brackets.
\item Additional controls: Journal, paper topic, and region of corresponding author. Full controls included, only significant controls displayed.
\end{tablenotes}
\end{threeparttable}
}
\end{table}

There are many considerations in this analysis, and in contemplating the
role of open science practices vis a vis researcher incentives. First,
researchers might not necessarily expect to see reduced review times
with open science practices. If a reviewer can follow and reproduce the
analysis in the paper during the review, this may result in more
thorough reviews and more extended resubmission processes. Improved and
more comprehensive reviews would benefit science, but would not
necessarily speed up the review process to the authors' benefit. Similarly, sharing data and code may
lead to more citations from others who pursue subsequent research, but
this effect is only likely if the original research question lends
itself to continued study, and if the code and data are packaged in such
a way that continued study is feasible. Further, the models presented in
Table~\ref{tbl-incentives} only explain one-tenth to one-fifth of the
variance in the data: important variables such as individual reviewer
availability, the complexity of the specific research question and
methods, and other sources of variance are not available to this
analysis but likely play an important role.

\section{Discussion}
\label{sec:discussion}

\subsection{Main findings}
The results of this study provide a field-level baseline for understanding open-science practices in transportation research. Rather than treating code and data availability as aggregate indicators alone, the analysis shows how these practices vary across publication venues, research areas, regions, and time. The findings allow identifying where open-science adoption is emerging, where barriers remain, and which parts of the research ecosystem may require targeted support.

\paragraph{Mean values} The code was available in 5\% of the papers. Data repositories were available in 4\% of the studies and open data were cited in 29\% of the studies. These results show that data sharing and code sharing are still rare in the field of transportation. These results are slightly higher than previous findings showing that fewer than 2\% of transportation simulation studies shared repositories with data and/or code~\citep{riehl2025revisiting}. The differences might be explained by the topics of the studies, the journals, the data extraction methods, and the availability definitions. The values are substantially lower than the rates reported in neighboring disciplines. For instance, previous studies found that code was available in 20\% of artificial intelligence conference papers in 2010-2019~\citep{lin2022automatic} and 18\% of agent-based modeling studies in 2018~\citep{janssen2020code}. 

\paragraph{Geographical effect}
The availability of code and data are strongly influenced by the location of the authors. Papers with authors based in Europe, South America, Africa, North America, and Oceania are more likely to have code available than those of authors in Asia. 
Papers with authors based in Europe, North America, Oceania and South America are more likely to have a data repository than those of authors in Asia, with Europe and North America showing the strongest effect. Possible explanations are differences in open science experience, practices, institutional valorization and funding policies. For instance, studies in other domains found that researchers in Asia were more interested in requiring permission to access their data~\citep{tenopir2015changes}. Researchers in low- and middle-income countries were more concerned about data misuse and privacy violations and had less experience with data sharing outside trusted collaborations~\citep{serwadda2018open}.

\paragraph{Temporal effect}
The availability of code and data are also strongly influenced by the year of publication. Newer papers more frequently provide code and data repositories, whereas older papers tend to reference existing datasets and do not provide a data repository. Possible explanations are changes in institutional and funding policies that favor open science ~\citep{nielsen2023using}, the increasing availability of data sharing infrastructure such as repositories and platforms, the introduction of a data availability statement in TR journals, and the promotion of open science principles by transportation associations in recent years ~\citep{caicedo2025sharing, wu2024reproducibility}. Studies in other domains have also shown that data and code sharing have gradually increased over time, influenced by journal policies ~\citep{roche2022paths, fink2025replication} and possibly technological advances ~\citep{fink2025replication}.

\paragraph{Influence of research topic}
The research topic significantly influences the availability of both code and data. Transportation research can be influenced by practices in other domains due to its multidisciplinary nature. Studies on data-driven modeling and prediction exhibit the highest levels of code and data availability. Researchers working on these topics may be influenced by practices that are well established in the machine learning and computer science community, where major conferences encourage the submission of data and code~\citep{pineau2021improving}. At these conferences, most authors are willing to submit code, a high number of reviewers are interested in consulting the code, and an increasing number of participants are joining reproducibility challenges ~\citep{pineau2021improving}. Studies on COVID-19 showed higher rates of citations to open data and data repositories than studies on optimization, but lower rates of code available. This result may be explained by the fact that several mobility datasets were released during the pandemic to support real time decision-making~\citep{combs2021shifting, li2021emerging}. In these emergency conditions, focusing on releasing data  may have been considered more valuable than code. Studies on urban environment showed higher rates of citations to open data and repositories than studies on optimization. These studies may use map and land use datasets that are publicly available or can be directly extracted from open source tools (e.g., OpenStreetMap). The barriers to data sharing may therefore be lower compared to other topics. Studies on traffic flow and transport emissions showed lower rates of code and data repositories, while studies on automated driving showed a low rate of code available and citations to existing data than studies on optimization. These topics are dominated by simulation studies in which the code may be difficult to share or tied to closed simulation tools. Citations to existing data on automated driving may be particularly limited because it is an emerging technology. These results are in alignment with previous findings showing low rates of code and data availability for simulation studies in transportation ~\citep{riehl2025revisiting}. The main reasons reported in surveys were high technical complexity, lack of resources, and lack of experience in sharing materials ~\citep{riehl2025revisiting}. Studies on driver behavior, transport policy, and travel mode showed low levels of code and data availability than studies on optimization. These transportation topics may rely on proprietary data and involve human participants, raising security, ethical, and privacy concerns. Key data in transportation modeling (e.g., operator data, GPS data and smart-card) are often commercial or inaccessible ~\citep{mahajan2022data}. The results also mirror patterns observed in the social sciences. A previous study found that willingness to share data is higher in hard sciences than in humanities and social sciences ~\citep{enwald2022data}. Studies on logistics, road infrastructure, supply chain showed the lowest levels of code and data availability. These topics may rely on simulations and proprietary data that agencies are less likely to share due to organizational protocols, security concerns, and lack of resources. A previous study on civil infrastructure engineering found that data of competitive edges, private data, and limited support for data sharing were among the main barriers to data sharing ~\citep{wang2023characterizing}. 

\paragraph{Influence of journal}
The journal paper significantly influences the availability of code and data. Even after controlling for the topic of the paper, differences in journal scope, editorial practices, cultural norms and target audience may influence these variations. Papers published in TR-C are the most likely to include data and code repositories. An explanation is that TR-C explicitly promotes open science initiatives and open datasets in its scope. Researchers who value open science may prefer to submit to this journal instead of others.
Papers published in TR-B are more likely to include code than the other journals (except TR-C) and less likely to cite open dataset than papers in TR-D and TR-A. These results may indicate that methodological papers value more their code as a contribution. These papers may use synthetic data for validation more often than citing open datasets. In contrast, papers published on TR-A and TR-D may use publicly available data more often due to the emphasis on policy-relevant contributions.
Papers published in TR-E are less likely to cite open datasets than papers published in TR-D and TR-A. The finding may be explained by the focus on applied logistics and supply chain, which was not fully captured by the topics. These studies often rely on proprietary data and simulations. 
Papers published in TR-IP are less likely to cite open data and have a data repository than papers in TR-D and TR-A. This result may be due to the different methodologies and data types that characterize interdisciplinary research and may hinder the release of open data.
Papers published in TR-F are the least likely to cite open data regardless of the data repository. This result may be due to the focus on human factors and traffic psychology, which is not fully captured by the topics. In this field, qualitative or sensitive data involving human subjects may not be publicly shared due to privacy concerns.

\paragraph{Publication characteristics}
The characteristics of the manuscript appear to influence open science practices, even after controlling for the topic and the journal. Papers with a high number of figures are more likely to have both code and data available. The high number of figures may be required to visualize computationally intensive workflows. Papers with a high number of tables are more likely to have citations to open data or a data repository. The high number of tables may be used to summarize quantitative datasets.
By contrast, papers with a high number of pages are less likely to have citations to open data and data repositories. This result may be linked to papers that make a more theoretical contribution without requiring extensive data and code.  Papers with a higher number of authors are less likely to have citations to open datasets. This result may reflect the reuse of data that are not publicly available with trusted collaborators.

\paragraph{The incentive gap}
Despite increasing momentum toward open science in the transportation research community, the findings point to a persistent misalignment in academic incentives. Within the observed period, there is no evidence that sharing code or data is associated with higher citation rates or decreased review times. Taken together, these patterns suggest that current peer-review and reward structures offer limited professional returns for the additional effort required to make research towards reproducible.

\subsection{Performance}
Based on our analysis, the performance of the extraction pipeline is sufficient for analysis. %
\sectionname~\ref{sec:agreement-analysis-results} shows that, for most features, the agreement between LLMs and the expert annotator is similar to the agreement between expert annotators. Metrics such as Fleiss's Kappa or Cohen's Kappa show that the LLM can achieve acceptable results on all features. We expect that the extraction and analysis pipeline can be readily adapted to other transportation journals. Given the relatively high agreement across the features considered, we additionally believe that numerous other questions about open science practices can be considered and validated using the same methodology.

\subsection{Limitations}

\subsubsection{Limitations in the sample for analysis}
The number of papers that were annotated by two experts was fairly small compared to the number of papers used to extract features. Because we used data up until 2025, our work is a snapshot of the state of availability until then. This can serve to monitor community progress, but these conclusions may not hold for future work.

\subsubsection{Limitations in the topic modeling}
The LDA used for topic modeling in this paper is also limited in its ability to classify papers. For instance, a number of papers classified under COVID-19 and pandemic were published before 2020 and refer to other emergency situations that limited mobility.  We would like to clarify that this comes from the nature of the LDA model: it is a generative probabilistic (Bayesian) model built on the bag-of-words assumption. Thus, it is an unsupervised probabilistic topic model that groups documents by word co-occurrence. Its output is a distribution over topics for each paper (each paper is a mixture of topics) and a distribution over words for each topic. We then name each topic grouping using the top words from each topic. Although LDA represents each paper as a mixture, we assign each paper to its single highest-probability (via \texttt{argmax}) topic. For interdisciplinary papers, which are common in transportation, this discards mixed-membership information and can place a paper in a category an expert may not choose first.

We explored other categorizations based on domain taxonomies, for example the Transportation Research Board (TRB) committee structure, but found it difficult to reach consistent agreement on how to assign papers under such schemes. We therefore adopted LDA, as it is the most widely used approach for topic analysis in the transportation research literature~\cite{sun2017discovering}. Topic categories enter the choice models as explanatory variables.  A different categorization could therefore change which topics reach significance and the magnitude of their coefficients. Our choice models do not use all 15 categories individually. As shown in Tables~\ref{tab:results_code} and~\ref{tab:results_data}, several topics turned out to have statistically similar effects, and these were therefore combined into grouped binary variables, yielding a coarser specification than the raw LDA output.

\subsubsection{Limitations in the LLM extraction}

While LLMs were the only solution to extract the necessary data with sufficient accuracy at the scale of the study, they are still imperfect. For example, LLMs could not perfectly recover metadata available in the XML file. Additionally, given that we did not set temperature to 0 during extraction, our exact output is difficult to reproduce. Given that human annotations are also not perfectly reproducible, this is a limitation inherent to studies of this kind. However, our pipeline and the extracted data are openly available.

\subsubsection{Limitations in the gap between availability and reproducibility}
While this study helps measure open science practices, the proposed pipeline does not actually measure paper reproducibility, which would require significant additional human labor. However, we can conclude that most transportation papers are currently very difficult to reproduce, given that they do not meet the necessary condition to share the data and code they used.

\section{Recommendations}
\label{sec:recommendations}
We offer evidence-based recommendations from the lens of open science––not as an ends in itself––but as an enabler for accelerating research progress and strengthening scientific integrity. Thus, our recommendations focus on pragmatism and aligning desired behavior with researcher incentives. Overall, because a researcher's primary incentive is to publish papers and garner citations, we believe that publishers have the highest leverage to influence open science practices in transportation, so we focus our recommendations there.

This study confirms that the transportation research community is in nascent stages of adopting open science practices. At the same time, it signifies that the community has a tremendous opportunity to shape the adoption of open science and the downstream impact. The following recommendations are informed by the findings and data produced from this study, the open science journeys of other research communities (both positive and cautionary), and by our collective decades of experience in the transportation research field. %

\subsection{Publishers}

For publishers, we recommend a balanced approach: promote practices that build momentum in open science, while avoiding passing an undue burden of their implementation solely onto paper authors. We thus recommend explicitly rewarding behaviors that make it easier for authors to engage in open science.

Our study found no significant difference in citation counts or review duration between papers that provided data and code and those that did not, suggesting that traditional academic metrics do not incentivize researchers to practice open science. This is in contrast to the sentiment expressed by some journals; for example, Transportation Research Part C already conveys a positive stance on open science, with their \href{https://www.sciencedirect.com/journal/transportation-research-part-c-emerging-technologies/publish/guide-for-authors}{guide for authors} stating that sharing data ``increases your exposure and may lead to new collaborations [and] provides readers with a better understanding of the described research''. While soft signaling led to significant impact in the machine learning community~\cite{pineau2021improving}, it appears insufficient for the transportation research community. At the most extreme, journals such as \href{https://pubsonline.informs.org/page/mnsc/code-and-data-disclosure-policy}{Management Science} require valid repository links (e.g., to GitHub, Zenodo, or Dataverse) or even full replicability prior to publication{~\cite{fivsar2024reproducibility,informs_ijoc_repositories,ferreira2010information}. We suggest doing so as a last resort\footnote{Compulsory open science creates additional challenges, such as inhibiting publication of research involving sensitive or proprietary data. Additionally, the machine learning community has demonstrated that compulsion is not necessary to achieve as high as 79\% code availability.}.

Rather than promote open science for the sake of open science, we suggest identifying policies or initiatives that better align researcher incentives with open science practices to accelerate transportation research. 
Success can be measured by highly and rapidly cited works that the community can build upon owing to its open data and code. At present, some of the most popular transportation datasets and benchmarks are not published in transportation journals~\cite{caesar2020nuscenes, sun2020scalability, ettinger2021large, barnes2020oxford, bock2020ind, wu2021flow, jiang2021dl}.
Thus, a high-leverage behavior that publishers can use to bootstrap the practice of open science in transportation is to encourage the publication of high-quality open datasets and benchmarks. Not only will this ease the open sourcing of downstream artifacts, it will create a virtuous cycle where community members have access to exemplars to develop more datasets and benchmarks. Open datasets and benchmarks have the added benefits of improving the comparability of research and thus measurement of progress, and attracting newcomers and talent from other fields. Journals can incentivize open datasets and benchmarks by hosting special issues~\cite{ETRR2025}, explicitly including open data and benchmarks within the aims and scope of the journal, and hiring dedicated editors to handle such topics.

Another strategy to give credit to researchers who contribute to open science is to augment the CRediT (Contributor Roles Taxonomy) author statement~\cite{allen2019can} to explicitly list open science practices---such as creating a data archive or reproducing a study---as contributions to a research project.

There are additional low hanging fruits that publishers can implement. For instance, we suggest mandating structured data and code availability statements in submissions. Currently, none of the journals considered encourages or requires a code availability statement. TR-A, TR-B, TR-D, TR-E, encourage authors to share a data availability statement, but do not require it, while TR-C, TR-F, and TRIP require authors to include a data availability statement. Compared to TR-D, only papers from TR-C were more likely to have a data repository regardless of data citations. Compared to papers published in TR-D, papers published in TR-B and TR-C are more likely to have a code repository available, while papers published in TR-A, TR-E, TR-F, and TR-IP were not significantly more likely to have a code repository available. In addition to encouraging open science practices, availability statements make it easier to measure the state of open science in the transportation community.

{As numerous communities have done,} peer-review guidelines could incorporate open science checklists without prolonging review times. {Furthermore, in the era of LLMs, publishers could implement automated checklists using LLMs to process the full text of a submitted paper, extending the process implemented in this paper. However, we caution against implementing these checklists without incorporating them into the reviewer and editor workflow, lest they turn into a box-ticking exercise with limited accountability---automated or not.} Additionally, drawing from successful initiatives in adjacent fields~\cite{acm-artifact-review-badging}, journals could award open science ``badges'' for verified availability, signaling quality and encouraging compliance. These policies should address transportation-specific challenges, such as large dataset sizes and regional data heterogeneity, by allowing exceptions for proprietary data while promoting anonymized alternatives.

\subsection{Conferences}

{Conferences can amplify the value of open science to the community. 
Conferences are the primary gathering place for researchers, and as such they play a critical role in conveying what is important and valued by the community. }
{Conferences have numerous mechanisms for directing focus and attention, including but not limited to keynote talks, awards, and tutorials. For example, the machine learning community successfully drew attention to a reproducibility crisis in reinforcement learning by dedicating a keynote~\cite{pineau2018reproducible} and several workshops to the topic.} {Similarly, conferences can issue paper awards associated with open science, such as ``Best Open Paper''. Conferences can signal the importance of sessions dedicated to open science by placing them in dedicated slots with no or few parallel sessions.}

\subsection{Funding agencies}

The absence of robust open science practices imposes hidden costs on funding agencies such as the US DOT, NSF, and EU Horizon programs, which collectively fund redundant data collection and methodological development across multiple research teams addressing similar questions. As such, funding agencies can intervene by prioritizing funding for research teams that commit to open sharing; for instance, in the United States, ``Funder requirement'' is a ``consistently leading motivating factor for data sharing'' \cite{Hahnel2024}. Funding agencies can also support the creation of datasets or benchmarks for topics with low adoption like supply chain logistics or resilience and evacuation, targeted incentives. {Discussed next, funding agencies can also help by supporting volunteer groups such as RERITE~\cite{rerite_website}, 
which commit to working across transportation research stakeholders to advance open science and scientific integrity.}

\subsection{Volunteer groups}
In some fields such as computer science and physics, grassroots volunteer-based efforts lead to development of open source tools and practices such as releasing preprints ~\cite{armeni2021towards}. However, these efforts face resistance in transportation, motivating dedicated open science communities like RERITE~\cite{rerite_website} that push the field at large forward to adopt these practices. {Volunteer groups can engage in scientific research to support evidence-based open science practices (such as this article), and can institutionalize the effort in the form of an annual ``transparency audit'' to support open science decision makers, such as publishers, conferences, authors, etc. More broadly, these volunteer groups can bridge the gap where existing institutions are not fully equipped to advance open science. Examples include: rewarding open science contributions by promoting and disseminating open research; aiding in the creation of pedagogical materials for dissemination within educational institutions, conferences, and summer schools\footnote{Practical skills include how to anonymize sensitive data, use version control for code, and cite shared resources properly. Example include the reproducibility tutorials at ITSC 2024 \cite{wu2024reproducibility} and ITSC 2025~\cite{rerite_tutorial_2025}. Our topic analysis and the extracted data availability statements could be mined to tailor educational material to specific community or sub-community needs.}; holding regular seminars and workshops to form an open science community within transportation; preparing reviewers, editors, and awards committees on how to evaluate open science practices; and, more generally, bootstrapping a culture of open science within the transportation community.}

\subsection{Research and educational institutions}

{Educational institutions can incorporate open science into their curricula. This can range from basic classroom modules on version control and data archiving to virtualization (e.g., Docker) and reproducibility workflows to intensive course projects that attempt to reproduce a transportation research paper.}

{Research institutions should recognize that research progress will be slow when researchers cannot build upon the work of others, as is the case when code and data availability are as low as they are currently in transportation research. They can explicitly factor this into hiring and promotion evaluations, and recognize the long-term scientific importance of dataset and benchmark contributions that lay the foundation for research progress.} Universities can also implement open science and data sharing policies, with many of the highest-publishing universities around the world already doing so \cite{Hahnel2024}.

\subsection{Authors}
{Authors can contribute by learning, implementing, and sharing open science practices. Authors should recognize that they are the primary beneficiaries of practicing open science---by making their code and data available to the community, they inadvertently make it available to \textit{themselves} and their collaborators six months down the line, when it comes time for a paper revision or a project extension. At the same time, it is more common that the implementation of open science is dictated by circumstance, for example based on the culture of the research group or the preferences of the research sponsor or data provider. Research group leaders can influence this by openly stating their support of open science.} Data containing proprietary or personal information may be difficult to share, but this is likely not the limiting factor for most of the papers that do not share data. The majority of data availability statements gave ``Data Available Upon Request'' as the reason the data was not shared, suggesting that privacy is not the primary reason and that a lack of training in how to share data likely plays a bigger role.

{We encourage authors to seek out creative ways to practice open science; for example, when working with proprietary data, authors can often make post-processed derived data available, or they can make their code available, along with some synthetic data. Even when open science is permitted, it is rarely encouraged, and there are numerous inhibitors along the way; for instance, some students have anecdotally told us that they sometimes feel embarrassed to make their code available because it's not ``pretty.'' While we insist that available is better than pretty, some authors' concerns will not be assuaged by this. For those authors, thankfully, today's LLMs can assist authors in writing documentation and README files, and cleaning up file and directory structure.}

\subsection{Researchers}
{We encourage researchers to continue conducting research to promote evidence-based decision making to advance transportation research. For example, interesting topics of further investigation could include more detailed analyses of transportation research topics that have notably high or low levels of open science practice, identifying and designing key missing benchmarks for transportation research, or exploring the use of LLMs for replicating key transportation research studies.
}

\section{Conclusions}
\label{sec:conclusion}
Our work highlights that the percentage of papers in transportation research using open science practices is critically low, with only 5\% sharing a code repository and 4\% sharing a data repository. Given the benefits of open science and the many other fields that have adopted such practices, it is both necessary and possible for the field of transportation research to improve. Given that there is not a significant link between open science practices and incentives like citations or reduced review time, encouraging more open science practices will likely require active changes made to be made by publishers, conferences, and funding bodies. Positive examples, such as TR-C, can serve as models for the rest of the community. Beyond the empirical results, this study provides a reusable measurement framework for transforming unstructured statements about research artifacts into structured and auditable open science indicators.

Furthering open science will also accelerate the field of transportation research itself. 
We found that 29\% of papers cite or link to an open dataset, yet only 4\% share a data repository. Given that some of these repositories contain repackaged datasets, less than 4\% of the papers we reviewed contributed new data to the transportation community. This severe asymmetry suggests that data from a few works form the basis of a large proportion of the field. If more authors released new open datasets, they could have a significant impact on the field. {We should also be aware of the risks of open science in the era of LLMs~\cite{acion2023generative}, that open science practices could facilitate the generation of fake research. However, we believe that the benefits of open science outweigh the risks, and that open science best practices can be designed to mitigate these risks. For example, sharing code and data can make it easier to detect and prevent fraud.}

Measuring open science is not a one-shot effort, which is why we have developed an automated pipeline that can continue to be used to track the progress of code and data availability in the transportation research journals. 

The framework developed in this study establishes a baseline and can be repeatedly applied to track changes in code and data availability in  journals and, with appropriate adaptation and validation, in other publication venues.
We are committed to transparency and reproducibility in our research, and we also benefit from the reproducible pipeline introduced in recent related work~\cite{riehl2025revisiting}, which enables our study to build upon prior efforts and move the field forward. We hope to periodically check new papers and release their statistics to monitor the community.

Additionally, there are a number of directions this work could be expanded. Although we did a limited exploration of the repositories we found, we hope to further study the documentation of the code and data shared. Additional exploration into funding bodies and open science practices could also yield insights about what policies are effective in encouraging open science.

\section*{Acknowledgment}
This research utilized data accessed through Elsevier’s APIs in compliance with Elsevier’s Text and Data Mining policy. 
We are grateful for the Reproducible Research in Transportation Engineering (\texttt{RERITE}) Working Group for providing us with the opportunity to build and strengthen this collaborative team. We appreciate the feedback and support from our colleagues, including Zuduo Zheng at University of Queensland, Christine Buisson at \'Ecole Nationale des Travaux Publics de l'\'Etat (ENTPE), Bidisha Ghosh at Trinity College Dublin, Joan Walker at UC Berkeley, Xinyu Chen at MIT, Gergely Zach\'ar, Daniel B. Work and Jonathan Sprinkle at Vanderbilt University, Micah Altman at MIT Libraries.
We gratefully acknowledge the volunteers who contributed to the manual validation dataset, including Khadidja Kadem, Cong Son Duong, Ali Naaman, Jincheng Ke, Maissa Mati, and Tuan Nguyen from Gustave Eiffel University COSYS/GRETTIA; Cameron Hickert, Jung-Hoon Cho, Shreyaa Raghavan, Vindula Jayawardana, Han Zheng, Hank Lin, Tianyue Zhou, and Edgar Ramirez Sanchez from MIT; Yiru Jiao from TU Delft; and Zhexian Li from USC. Their efforts were essential to the accuracy and reliability of our feature extraction process.
Early versions of this work were presented at Transportation Research Symposium (TRS) 2025, Transportation Research Board Conference on Data and AI for Transportation Advancement (TRB DATA) 2025, International Symposium on Transportation Data and Modeling (ISTDM) 2025, Modeling Mobility Conference (MoMo) 2025, and ITS Irvine Symposium on Emerging Research in Transportation 2026. We appreciate the feedback we received from these presentations. Given her role as Editor, Cathy Wu had no involvement in the peer-review of this article and has no access to information regarding its peer-review. Full responsibility for the editorial process for this article was delegated to another journal editor. We thank the anonymous reviewers for their constructive comments, which have helped improve this work.

This research was partially supported by the National Science Foundation (NSF) CAREER award \#2239566 (Wu), NSF award \#2434399 (Wu), the Marie Sklodowska-Curie COFUND grant agreement no. 101034248 (Belkessa, Ameli), French ANR award \#ANR-24-CE22-7264 (Ameli) and \#ANR-24-CE22-2750 (Varotto). The views expressed herein are those of the authors and do not necessarily reflect the views of the funding agencies.

\section*{Declaration of generative AI use}
During the preparation of this work, the author used Google Gemini 2.5 in order to extract open science features from scientific papers. After using this tool/service, the authors reviewed and edited the content as needed and take full responsibility for the content of the published article.

\section*{Data availability}
We are committed to transparency and reproducibility in our research. The data used in this study include (\romannumeral1) raw full-text articles from Elsevier Transportation Research journals, (\romannumeral2) features extracted using large language models (LLMs), and (\romannumeral3) a manual labeled dataset for validation.
The raw dataset for the full-text paper is not available due to the copyright policy of Elsevier~\cite{ElsevierTDMlicense} for text and data mining, but the pipeline to extract the dataset using the API is available in our code repository.
Our code is available at \url{https://anonymous.4open.science/r/most-pipeline/} for peer review.

\appendix
\section{Feature definitions}
\label{appendix:definition}

These were the feature definitions provided to manual annotators.

\begin{description}

    \item[\texttt{is\_quantitative\_study}:] Did the paper perform any quantitative calculations?

    \item[\texttt{is\_code\_publicly\_available}:] Does the paper claim to share the code used?
    \begin{itemize}[noitemsep, parsep=1pt, topsep=1pt, partopsep=0pt]
        \item \textbf{True} only if there is a link to a website, repository, or files containing code for the project.
        \item The link does not have to be active, it just needs to be stated in the paper. (i.e. “We share our code at nolongeractivelink.org.” - link activeness will be checked by another feature)
    \end{itemize}

    \item[\texttt{is\_code\_link\_valid}:] Is the link actually live and contains code files?

    \item[\texttt{is\_data\_repository\_available}:] Does the paper share a data repository? The repository must be created by the authors but does not need to contain data the authors created; i.e. it can contain processed data from NGSIM.
    \begin{itemize}[noitemsep, parsep=1pt, topsep=1pt, partopsep=0pt]
        \item \textbf{True} if the paper shares a link containing all data used for the experiment.
        \item \textbf{True} if a paper used a simulated dataset and shares code to fully reproduce the data.
        \item \textbf{False} if only certain sample instances of data are shared.
        \item \textbf{False} if the paper used data from NGSIM and links to NGSIM, but does not link to a repository where their preprocessed data is stored.
        \item \textbf{False} if the only data available is through supplementary materials (not accessible by everyone).
    \end{itemize}

    \item[\texttt{is\_data\_link\_valid}:] Is the link actually live and contains data files?

    \item[\texttt{is\_data\_cited}:] Does the paper use and cite at least 1 publicly available dataset? (This would include datasets released by governments as well as datasets like NGSIM, Sioux Falls, and Waymo dataset.)
    \begin{itemize}[noitemsep, parsep=1pt, topsep=1pt, partopsep=0pt]
    \item \textbf{True} if a paper uses Open Street Map and a private dataset from an e-bike company.
    \item \textbf{True} if email/phone registration is required, but False if requires institutional email/membership or payment.
    \item \textbf{False} if a paper uses a survey they conducted and private data from an airline.
    \item \textbf{False} if a paper uses data from a nonprofit or government that is not publicly available online.
    \end{itemize}

\end{description}

\section{Prompts used for LLM feature extraction}
\label{appendix:prompt}

\begin{promptcard}{Extracting code-related features}
 You are an expert assistant specializing in academic paper computational reproducibility analysis. Your task is to extract specific metadata about availability from research papers. You must respond with only a valid \texttt{JSON} object, without any additional text comments, or markdown formatting.
The papers are from Transportation Research journals.
The JSON object must conform to the following schema:
\begin{lstlisting}[breaklines=true, basicstyle=\ttfamily]
{
  "title": "string", // The exact title of the paper.
  "is_code_used": "boolean", // Did the paper use code to reach its conclusions?
  "reason_code_is_used": "string", // **If code was used, you MUST extract the justification directly from the paper. This field MUST contain a VERBATIM quote. Do not summarize or paraphrase.**
  "reason_code_is_not_used": "string", // **If code was not used, you MUST extract the justification directly from the paper. This field MUST contain a VERBATIM quote. Do not summarize or paraphrase.**
  "is_code_publicly_available": "boolean", // If code was used, is is publicly available?
  "reason_code_available": "string", // **If code is available, you MUST extract the justification directly from the paper. This field MUST contain a VERBATIM quote. Do not summarize or paraphrase. For example, if the paper says, \'The code for this project can be found on GitHub\', you must return that exact phrase. If no reason is stated, return an empty string.**
  "reason_code_unavailable": "string", // **If code is not available, you MUST extract the reason directly from the paper. If no reason iss tated, return an empty string. This field MUST contain a VERBATIM quote. Do not summarize or paraphrase.**
}
...
\end{lstlisting}
\end{promptcard}

\begin{promptcard} {Extracting data-related features}
You are an expert assistant tasked with extracting metadata about research dataset availability — defined here as processed data by the team — from papers published in Transportation Research journals. The journal policy encourages research data deposit, citation and linking. Which means that there are 3 categories of data availability: (i) data repository (e.g. Zenodo, Figshare, etc.), (ii) data cited or linked to a public source (e.g. open data from a government agency or other research institutions), and (iii) data not available.

Reply with one valid JSON object only - no extra text, comments, or markdown—using the schema below: 

\begin{lstlisting}[breaklines=true, basicstyle=\ttfamily]
    { "title": "string", // Exact paper title from the markdown file title

"is_data_used": "boolean", // Did the study explicitly rely on a dataset (not just a source/statistics from a report) to produce results (e.g. a paragraph describing the data used)?

"is_simulation_study": "boolean", // Is the study a simulation study (e.g. purely using a simulation tool to produce results)?

"reason_simulation_study": "string", // If yes, VERBATIM justification from the paper (no paraphrasing)

"reason_data_is_used": "string", // If yes, VERBATIM justification from the paper (no paraphrasing)

"is_data_repository_available": "boolean", // Is the data available in a repository (e.g. Zenodo, Figshare, Github, etc.)? Data is NOT the same as code or implementation or algorithms, but both can be in the same repository.

"reason_data_repository_available": "string", // If yes, VERBATIM justification from the paper (no paraphrasing)

}
...
\end{lstlisting}
\end{promptcard}
\textit{Note: The rest of the prompts can be found on our repository.}

\pagebreak

\section{Justification for the usage of LLMs}
\label{appendix:just}

To justify our use of LLMs, we compare our process to the method developed and outlined by \cite{riehl2025revisiting}, which uses text search (TS) to find relevant links and keywords in the text.

\begin{table}[htbp]
\caption{Kappa values for text matching feature extraction versus LLM extraction. Interpretation guide with Fleiss's $\kappa$: 0.21-0.40 (\colorbox{red!15}{Fair}), 0.41-0.60 (\colorbox{orange!25}{Moderate}), 0.61-0.80 (\colorbox{yellow!40}{Substantial}), and 0.81-1.00 (\colorbox{green!20}{Almost Perfect})~\cite{landis1977}.}\label{kappa-ts}
\centering
\footnotesize
\begin{tabular}{rccccc}
\toprule
\textbf{Feature} & \textbf{\makecell{Cohen Kappa\\(H1 vs H2)}} & \textbf{\makecell{Cohen Kappa\\(H1 vs TS)}} & \textbf{\makecell{Cohen Kappa\\(H2 vs TS)}} & \textbf{\makecell{Fleiss Kappa\\(H1 vs H2 vs TS)}} & \textbf{\makecell{Fleiss Kappa\\(H1 vs H2 vs LLM)}} \\
\midrule
\texttt{is\_quantitative\_study}               & 0.4757 & -0.0365 & -0.0345 & 0.1782 & 0.4586 \\
\texttt{is\_code\_publicly\_available} & 0.7534 & 0.5922 & 0.5961 & 0.6364 & 0.8388 \\
\texttt{is\_data\_repository\_available}               & 0.3333 & 0.3592 & 0.0986 & 0.2618 & 0.3994 \\
\texttt{is\_data\_cited}               & 0.6672 & 0.0332 & 0.0275 & 0.0360 & 0.5224 \\
\bottomrule
\end{tabular}
\end{table}

\begin{table}[htbp]
\caption{Agreement percentages for text matching feature extraction versus LLM.} \label{pa-ts}
\centering
\footnotesize
\begin{tabular}{rccccc}
\toprule
\textbf{Feature} & \textbf{\makecell{Percentage\\Agreement\\(H1 vs H2)}} & \textbf{\makecell{Percentage\\Agreement\\(H1 vs TS)}} & \textbf{\makecell{Percentage\\Agreement\\(H2 vs TS)}} & \textbf{\makecell{Percentage\\Agreement\\(H1 vs H2 vs TS)}} & \textbf{\makecell{Percentage\\Agreement\\(H1 vs H2 vs LLM)}} \\
\midrule
\texttt{is\_quantitative\_study}       & 0.9062                          & 0.8646                            & 0.8958                           & 0.8333                                   & 0.8854    \\
\texttt{is\_code\_publicly\_available} & 0.9688                          & 0.9271                            & 0.9375                            & 0.9167                                   & 0.9688                                  \\
\texttt{is\_data\_repository\_available}               & 0.9271                          & 0.8854                            & 0.8750                            & 0.8438                                   & 0.9167                                 \\
\texttt{is\_data\_cited}       & 0.8542                          & 0.3750                            & 0.3333                            & 0.2812                                   & 0.6979    \\
\bottomrule
\end{tabular}
\end{table}

\begin{table}[htbp]
\caption{Prevalence of each feature found by H1, H2 and text search (TS).}
\label{prev-ts}
\centering
\footnotesize
\begin{tabular}{rccc}
\toprule
\textbf{Features}            & \textbf{Prevalence (H1)} & \textbf{Prevalence (H2)} & \textbf{Prevalence (TS)} \\
\midrule
\texttt{is\_quantitative\_study}        & 0.8854                   & 0.9167                   & 0.9792                \\ 
\texttt{is\_code\_publicly\_available} & 0.0833                   & 0.0521                   & 0.1146                   \\
\texttt{is\_data\_repository\_available}               & 0.0833                   & 0.0312                   & 0.1146                   \\
\texttt{is\_data\_cited}        & 0.3438                  & 0.3021                   & 0.9688                 \\ 
\bottomrule
\end{tabular}
\end{table}

The agreement percentage and Fleiss Kappa of text search is lower for all of the features because text search overestimated the prevalence (Table \ref{prev-ts}). Most notably, \texttt{is\_code\_publicly\_available} has moderate agreement (0.6364) using text search, while having substantial agreement with an LLM (0.8388). For \texttt{is\_code\_publicly\_available}, we find that a simple text search for a github repository link overestimates code availability, leading to its decreased accuracy (Table \ref{prev-ts}). Finally, text search cannot separate data and code repositories.

\section{Data postprocessing}
\label{appendix:postprocess}
Postprocessing is fully deterministic and is designed to (i) separate paper-level attributes from artifact-level evidence and (ii) ensure that all downstream variables are computed from a consistent, auditable set of intermediate tables. We construct three relational tables keyed by DOI: a \emph{paper-level} table (one row per DOI) containing bibliographic metadata, topic and region assignments, and final availability indicators; an \emph{artifact-level} table (one row per extracted dataset or artifact associated with a DOI) containing repository attributes such as host category and link validity; and a \emph{link-level} table (one row per DOI--URL pair) used to canonize and deduplicate URLs and to compute per-paper counts by host category. The main steps are as follows:

\begin{enumerate}[label=(\roman*),noitemsep,parsep=1pt,topsep=1pt,partopsep=0pt]
    \item \textbf{Parsing and flattening.} We parse the Elsevier \texttt{XML} metadata and merge them with the LLM \texttt{JSON} outputs. Nested structures (e.g., multiple dataset mentions or multiple links per paper) are flattened into the artifact- and link-level tables, while document-level fields (e.g., journal, year, authorship, and topic assignment) are stored in the paper-level table keyed by DOI.

    \item \textbf{Identifier and type normalization.} We standardize identifiers and field types across sources. DOIs are lowercased, stripped of common prefixes (e.g., \texttt{doi:}), and checked for format validity to preserve referential integrity across the relational tables. Boolean fields returned by the LLM (e.g., \texttt{is\_code\_publicly\_available} and \texttt{is\_data\_repository\_available}) are coerced to consistent logical values; when the manuscript does not provide sufficient information, values are retained as missing rather than imputed.

    \item \textbf{URL canonization, classification, and validation.} We collect all outbound URLs extracted from full text and LLM outputs, deduplicate them per DOI, and canonize them to a consistent form (e.g., standardizing schemes and removing parameters that do not change resource identity). Each URL is classified into host categories based on its domain (e.g., collaborative code hosts; DOI-minting general-purpose repositories; institutional repositories; and government open-data portals). We then perform automated validity checks to determine whether links are live and whether the landing content corresponds to the claimed artifact type (code versus data). Validity indicators are retained at the link/artifact level and later aggregated to paper-level variables.

    \item \textbf{Aggregation and feature construction.} Artifact- and link-level indicators are aggregated to paper-level variables used in \sectionname~\ref{sec:analysis}. In particular, we derive binary indicators for whether a paper provides any code link, whether a data repository is provided, whether at least one dataset is cited or linked, and whether at least one artifact link is valid. We also derive contextual covariates used in modeling (e.g., paper age, journal part, LDA topic, and corresponding-author region), along with summary link-count measures by host category.

    \item \textbf{Integrity checks and dataset filtering.} We apply consistency checks across tables, including DOI uniqueness, one-to-many coherence between paper- and link-level records, and logical consistency between availability flags and the presence of at least one URL in the corresponding host category. Records with missing or invalid DOIs, or with irreconcilable inconsistencies, are excluded. After these checks, 10{,}480 papers are retained from the 10{,}724 full-text articles.
\end{enumerate}

Figure~\ref{fig:postprocessing} illustrates the postprocessing pipeline as a flowchart.
\begin{figure}[htpb]
    \centering
    \includegraphics[width=\textwidth]{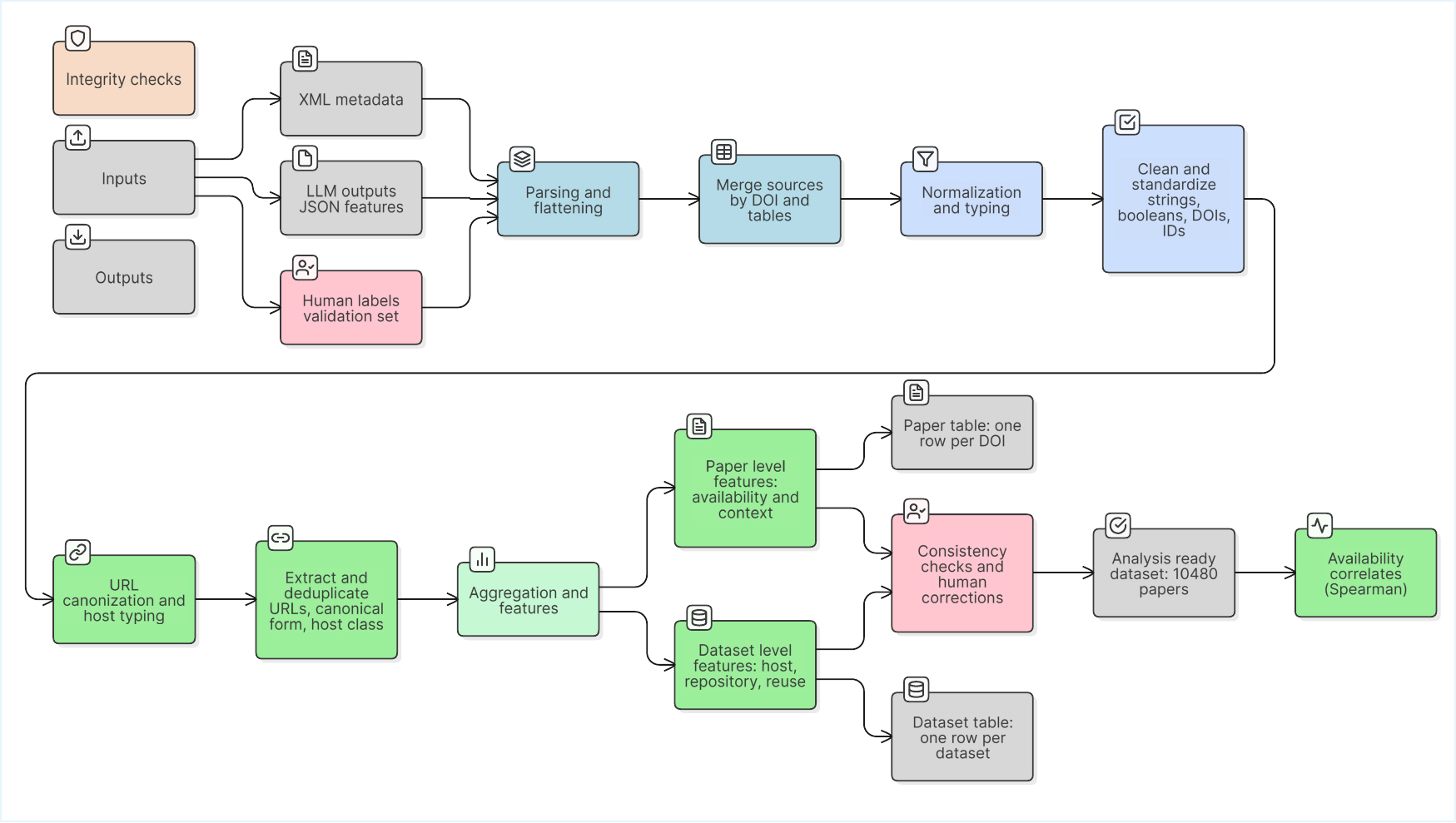}
    \caption{Flowchart of the postprocessing pipeline. Starting from raw XML metadata, LLM-extracted \texttt{JSON} features, and human annotations, the pipeline (i) parses and flattens inputs into paper- and dataset-level tables, (ii) normalizes types and encodings, (iii) canonizes and classifies URLs into code and data host categories, and (iv) aggregates and validates features to produce the final analysis-ready dataset.}
    \label{fig:postprocessing}
\end{figure}

\section{Data dictionary and features}
\label{appendix:data_dict}

In addition to the metadata features, LDA extracted topics, and LLM extracted features, we also derived additional features. A data dictionary of all features in the final dataset are listed below with their type, origin, and description.
\begin{table}[htbp]
\centering
\scriptsize
\setlength{\tabcolsep}{4pt}
\renewcommand{\arraystretch}{1.05}
\caption{Post-processed and transformed features obtained from initial features in \tablename~\ref{appendix:definition}.}
\label{tab:data-dict}
\begin{tabularx}{\textwidth}{P{3.4cm} P{1.6cm} P{2.6cm} X}
\toprule
\textbf{Name} & \textbf{Type} & \textbf{Stage / Origin} & \textbf{Description} \\
\midrule
code\_link & object & parsing & Raw link to code, unprocessed or malformed \\
code\_link\_by\_the\_author & object & LLM extraction & Whether the code link was explicitly provided by the authors \\
programming\_language & object & metadata parsing & Named programming language(s) used in the study (free text) \\
software\_used & object & metadata parsing & Software tools or packages declared in method or appendix \\
cpu\_used & object & metadata parsing & CPU model or specifications when computational config is reported \\
gpu\_used & object & metadata parsing & GPU type used when deep learning or simulations are involved \\
os\_used & object & metadata parsing & Declared operating system used in experiments \\
ram\_used & object & metadata parsing & Amount of RAM reported (in GB) \\
lda\_topic & int64 & topic modeling & Raw LDA-inferred topic index before mapping to label \\
clean\_primary\_region & object & geographic parsing & Cleaned world region label based on author affiliation \\
is\_data\_cited\_or\_linked & bool & LLM extraction & Whether the data was explicitly cited or linked in the reference list \\
is\_simulation\_study & bool & LLM extraction & Flag indicating use of simulated or synthetic data in the analysis \\
links\_to\_the\_data\_repository & object & parsing & Raw list of extracted links to dataset repositories \\
primary\_continent & object & geographic parsing & Redundant or preprocessed version of affiliation continent \\
n\_open\_repo & int64 & aggregation & Count of dataset links hosted in open-science repositories (Zenodo, Figshare, etc.) \\
n\_code\_links & int64 & aggregation & Total number of code links parsed (regardless of validity) \\
has\_code\_link\_gitlab & bool & derived feature & Whether a GitLab URL is included in the code links \\
has\_code\_link\_bitbucket & bool & derived feature & Whether a Bitbucket URL is included in the code links \\
availability\_reasons & object & LLM extraction & LLM-provided rationale for dataset/code availability or absence \\
n\_src\_benchmark\_hub & int64 & source counts & Number of links coming from benchmark-hosting hubs (e.g., PapersWithCode) \\
n\_src\_code\_host & int64 & source counts & Number of code links from recognized code hosting platforms \\
n\_src\_gov\_open\_data & int64 & source counts & Number of dataset links pointing to government open data portals \\
n\_src\_ngo\_nonprofit\_org & int64 & source counts & Links to NGO or nonprofit-hosted datasets \\
n\_src\_open\_repo\_doi & int64 & source counts & Number of datasets hosted on DOI-backed repositories (e.g., Zenodo) \\
n\_src\_other\_web & int64 & source counts & Links to generic web URLs not falling in known host categories \\
n\_src\_university\_repo & int64 & source counts & Number of links to university-hosted data repositories \\
uses\_data & bool & derived feature & Final flag indicating actual dataset use after manual correction \\
paper\_age\_years & float64 & derived metric & Years since publication (based on system crawl date) \\
region\_normalized & object & derived feature & Encoded world region label after mapping \\
is\_region\_asia & bool & one-hot encoding & Whether primary affiliation is located in Asia \\
is\_region\_europe & bool & one-hot encoding & Whether primary affiliation is located in Europe \\
is\_region\_north\_america & bool & one-hot encoding & Whether primary affiliation is located in North America \\
is\_region\_oceania & bool & one-hot encoding & Whether primary affiliation is located in Oceania \\
is\_region\_south\_america & bool & one-hot encoding & Whether primary affiliation is located in South America \\
is\_region\_africa & bool & one-hot encoding & Whether primary affiliation is located in Africa \\
n\_regions\_listed & int64 & derived feature & Number of distinct global regions covered by author affiliations \\
journal\_rate\_atleast & float64 & journal aggregation & Journal-level share of papers with at least one dataset available \\
journal\_rate\_all & float64 & journal aggregation & Journal-level share of papers with all datasets available \\
journal\_n & int64 & journal aggregation & Number of papers published in the journal in the dataset \\
papers\_is\_data\_used\_true & int64 & journal aggregation & Raw count of papers that used data \\
papers\_is\_data\_used\_share & float64 & journal aggregation & Share of papers in journal that used data \\
papers\_is\_data\_cited\_or \_linked\_share & float64 & journal aggregation & Share of papers linking or citing datasets \\
papers\_atleast\_one\_dataset \_available\_share & float64 & journal aggregation & Share of papers with at least one dataset accessible \\
papers\_all\_datasets \_available\_share & float64 & journal aggregation & Share of papers with all datasets accessible \\
papers\_uses\_data\_share & float64 & journal aggregation & Share of papers in journal that used data (final label) \\
\bottomrule
\end{tabularx}
\end{table}

\section{Topic}
\label{appendix:topic}
Topics were extracted using the gensim \cite{rehurek_lrec} package's LDA with 15 categories (decided by using coherence, with 15 categories having a coherence of 0.499). The LDA model was given the title, abstract, journal name, and keywords. The preprocessing step removed all non-alphabetic characters, excess spaces, emails, and made everything lowercase.
The most relevant words for each topic are listed in \figurename~\ref{fig:main-grid}.

\begin{figure}[htpb]
    \centering

    \begin{subfigure}[t]{0.32\textwidth}
        \centering
        \includegraphics[width=0.8\textwidth]{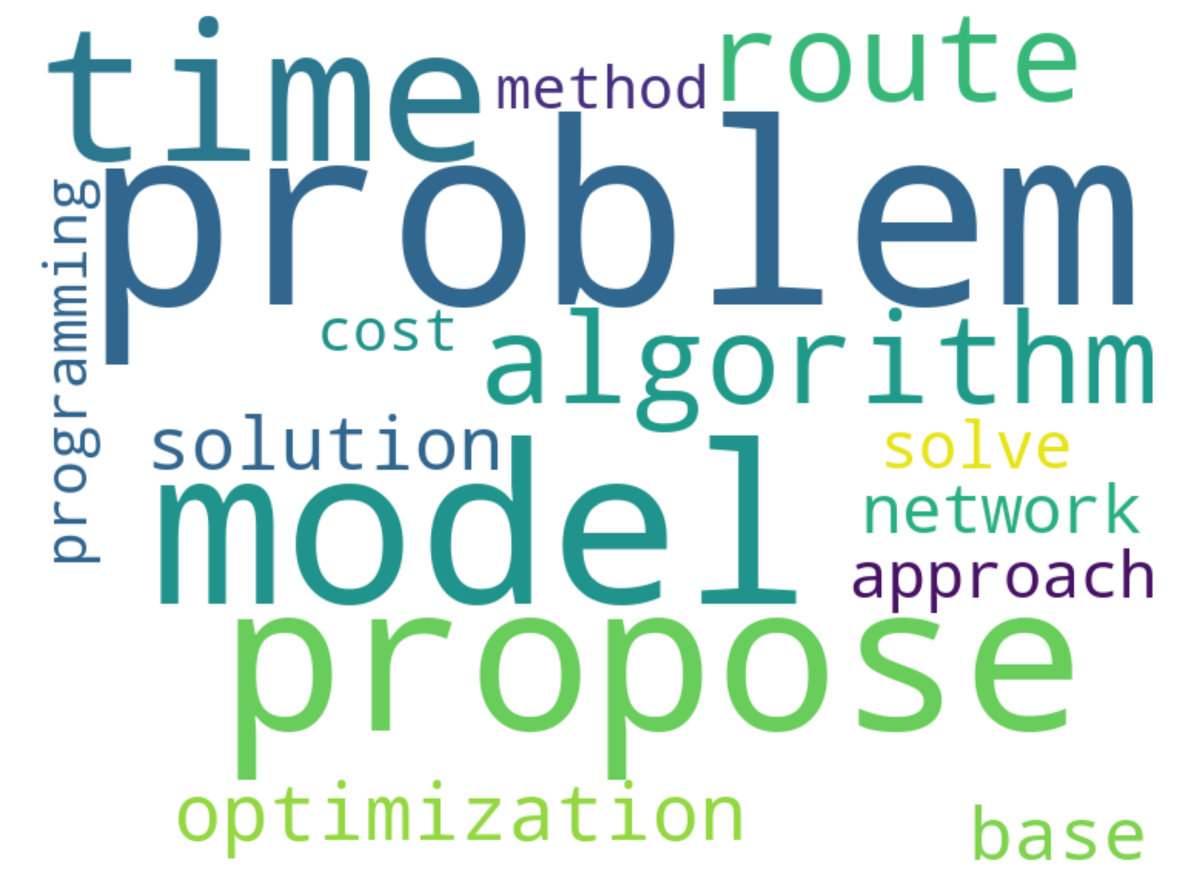}
        \caption{Optimization}
        \label{fig:opt-routing}
    \end{subfigure}
    \hfill %
    \begin{subfigure}[t]{0.32\textwidth}
        \centering
        \includegraphics[width=0.8\textwidth]{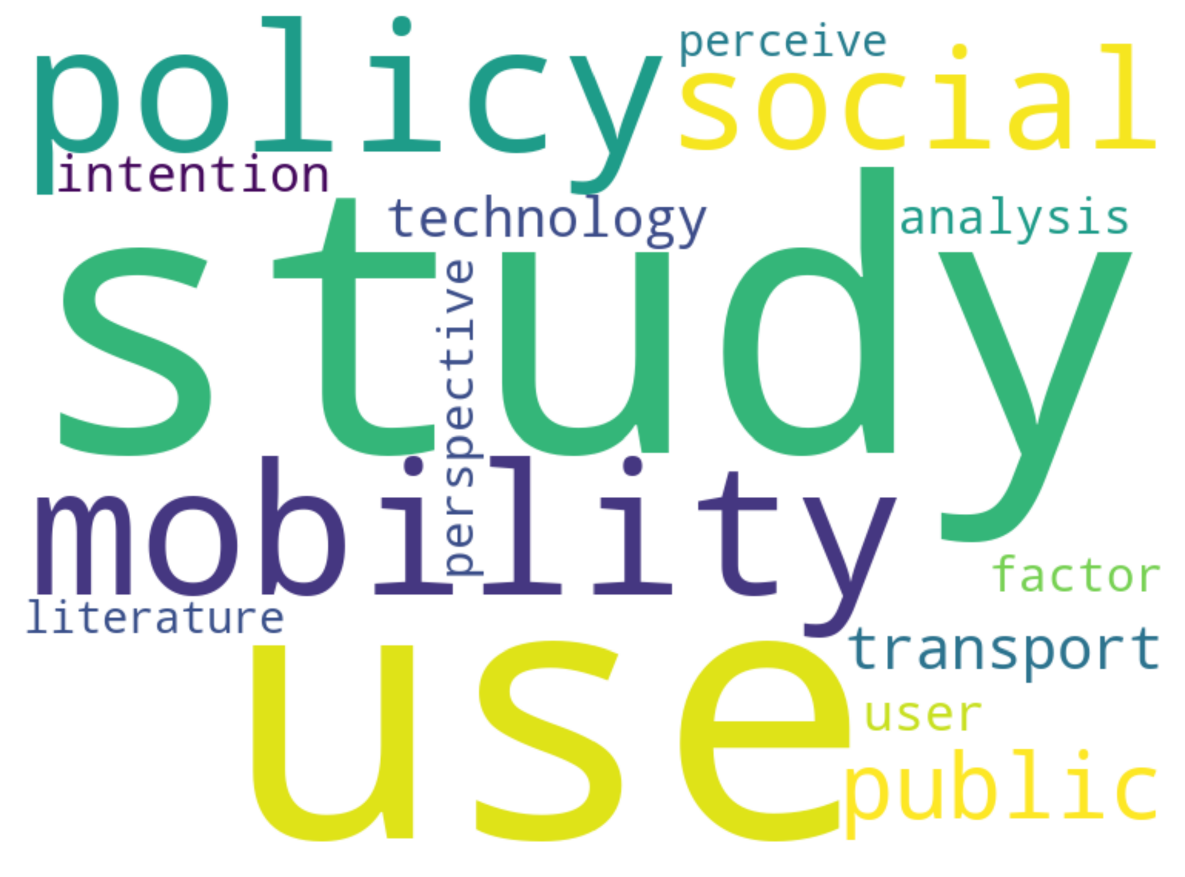}
        \caption{Mobility policy}
        \label{fig:social-policy}
    \end{subfigure}
    \hfill
    \begin{subfigure}[t]{0.32\textwidth}
        \centering
        \includegraphics[width=0.8\textwidth]{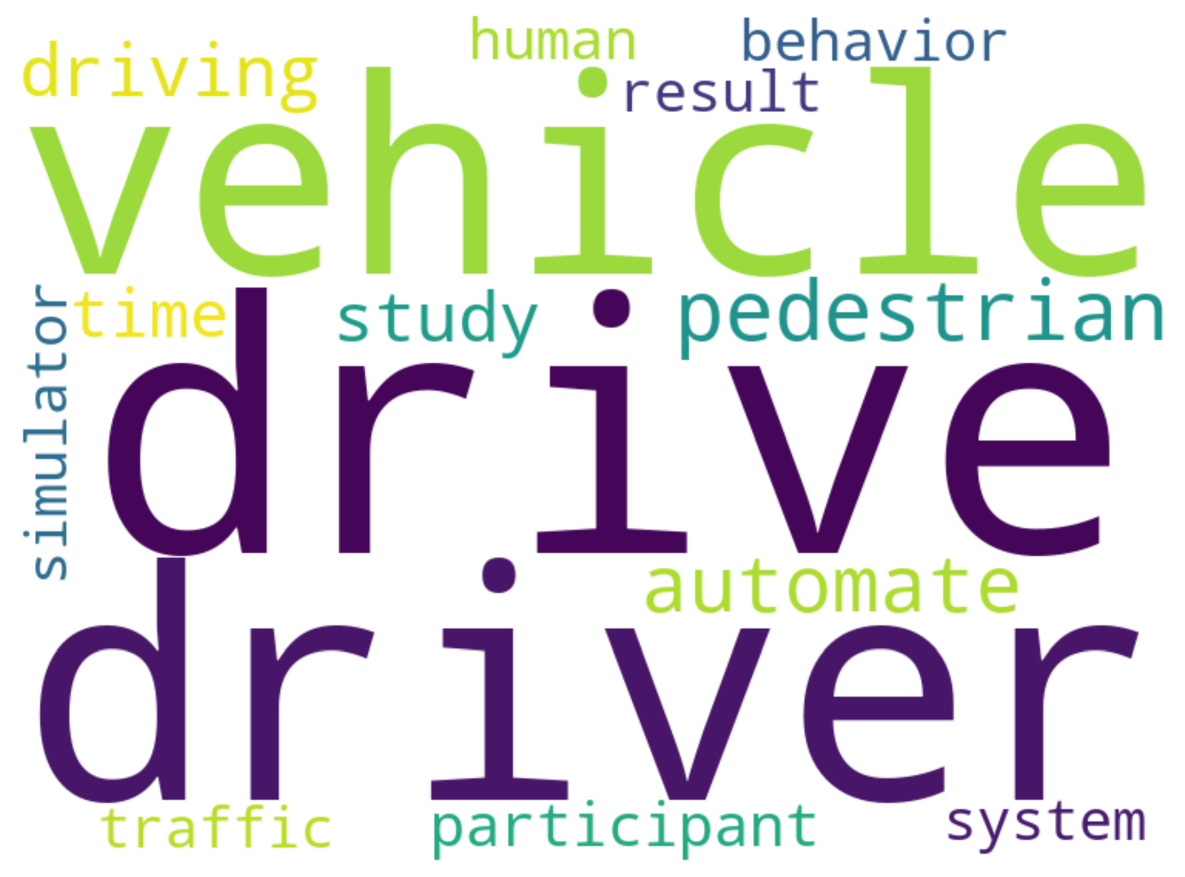}
        \caption{Driving behavior in automated driving}
        \label{fig:emerging-impacts}
    \end{subfigure}
    
    \vspace{0.5em}

    \begin{subfigure}[t]{0.32\textwidth}
        \centering
        \includegraphics[width=0.8\textwidth]{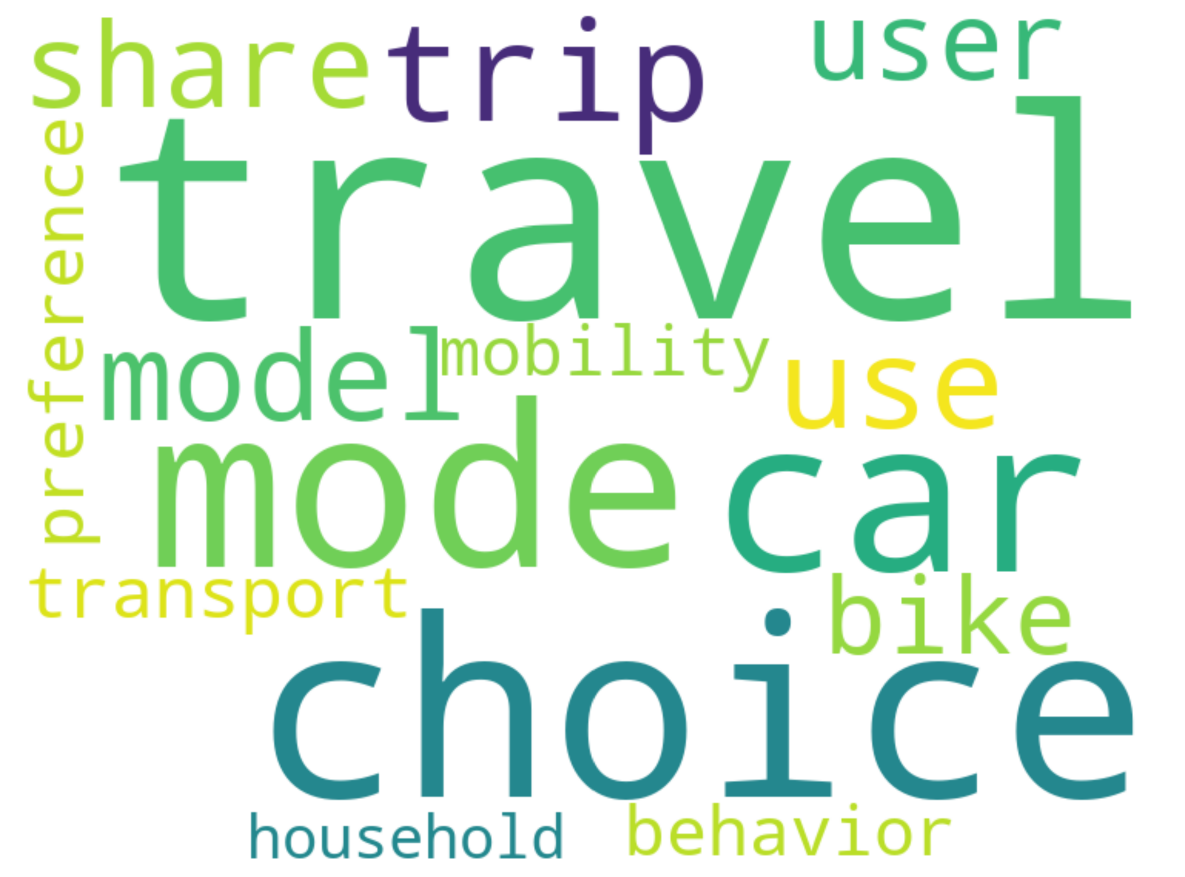}
        \caption{Choice model}
        \label{fig:public-transit}
    \end{subfigure}
    \hfill
    \begin{subfigure}[t]{0.32\textwidth}
        \centering
        \includegraphics[width=0.8\textwidth]{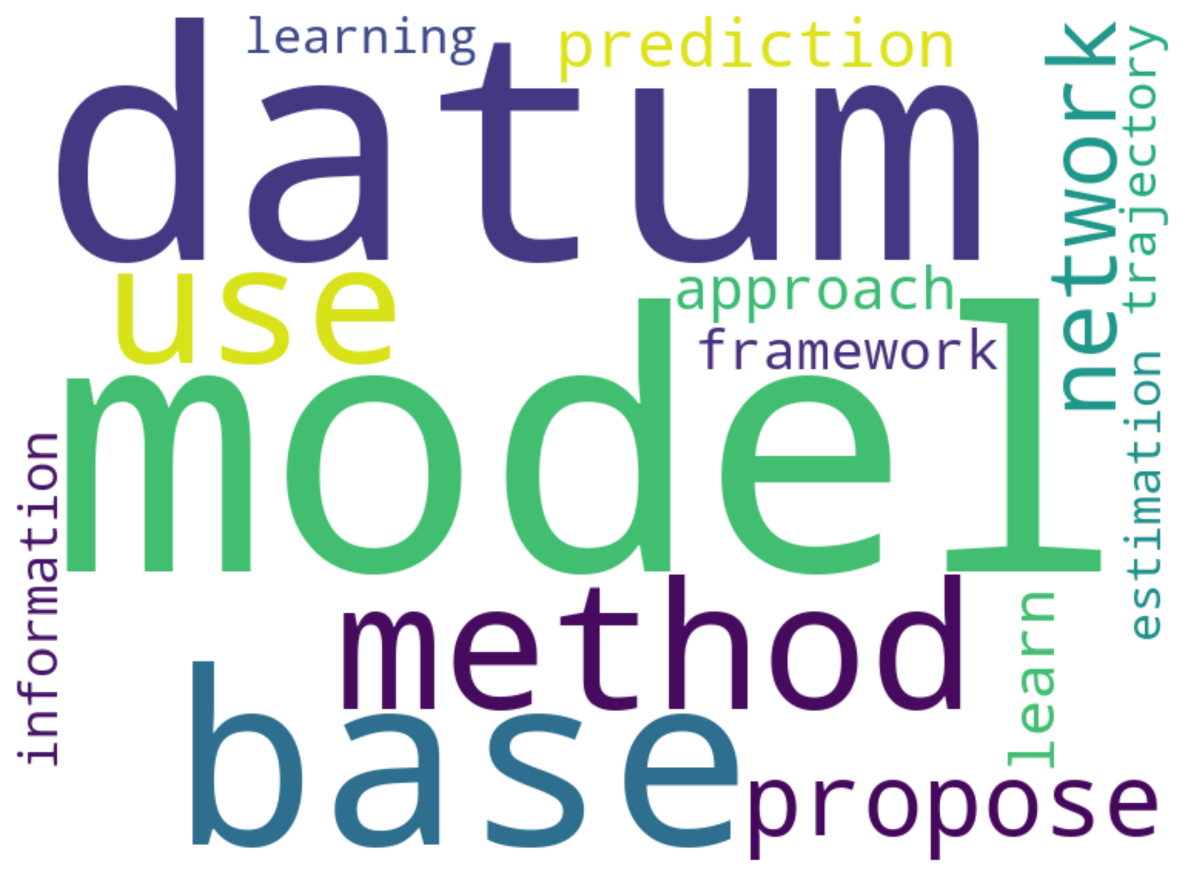}
        \caption{Data-driven modeling and prediction}
        \label{fig:road-infra}
    \end{subfigure}
    \hfill
    \begin{subfigure}[t]{0.32\textwidth}
        \centering
        \includegraphics[width=0.8\textwidth]{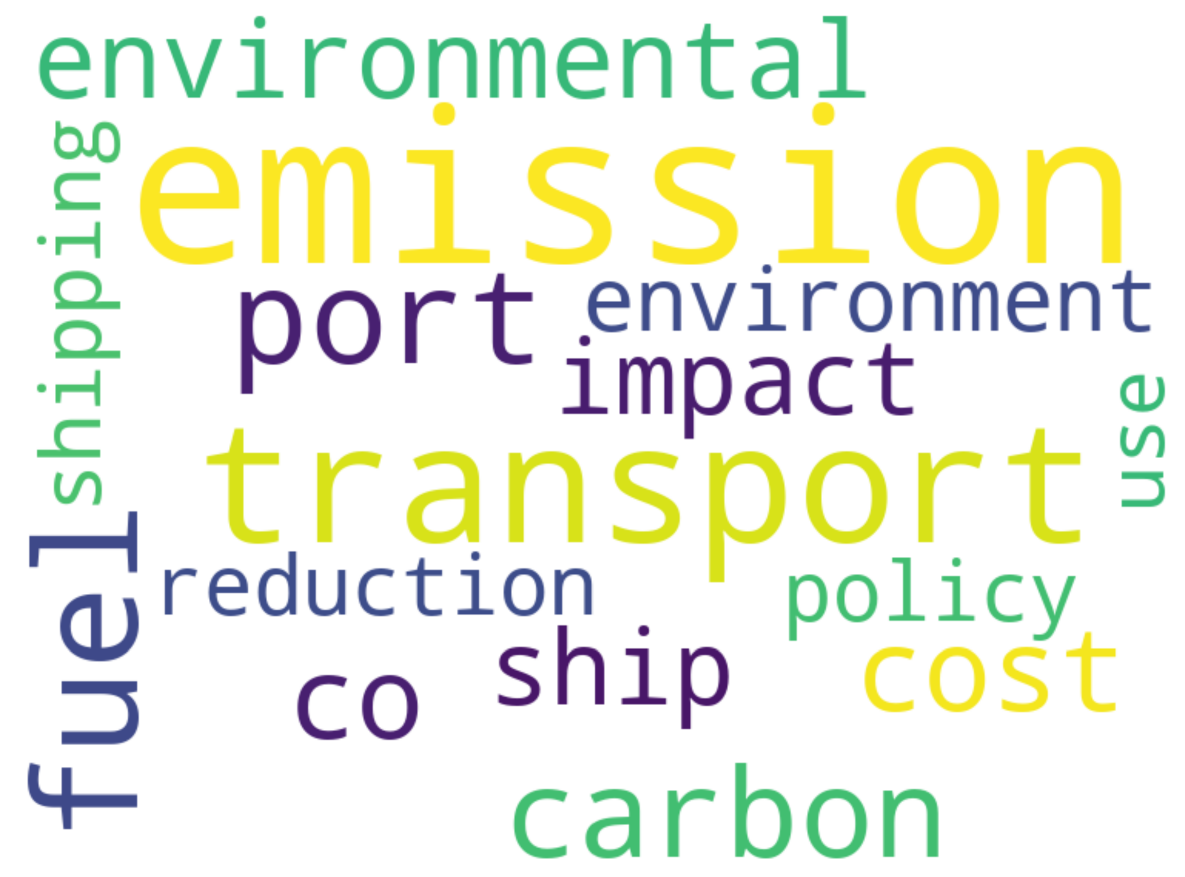}
        \caption{Emissions}
        \label{fig:supply-chain}
    \end{subfigure}

    \vspace{0.5em}

        \begin{subfigure}[t]{0.32\textwidth}
        \centering
        \includegraphics[width=0.8\textwidth]{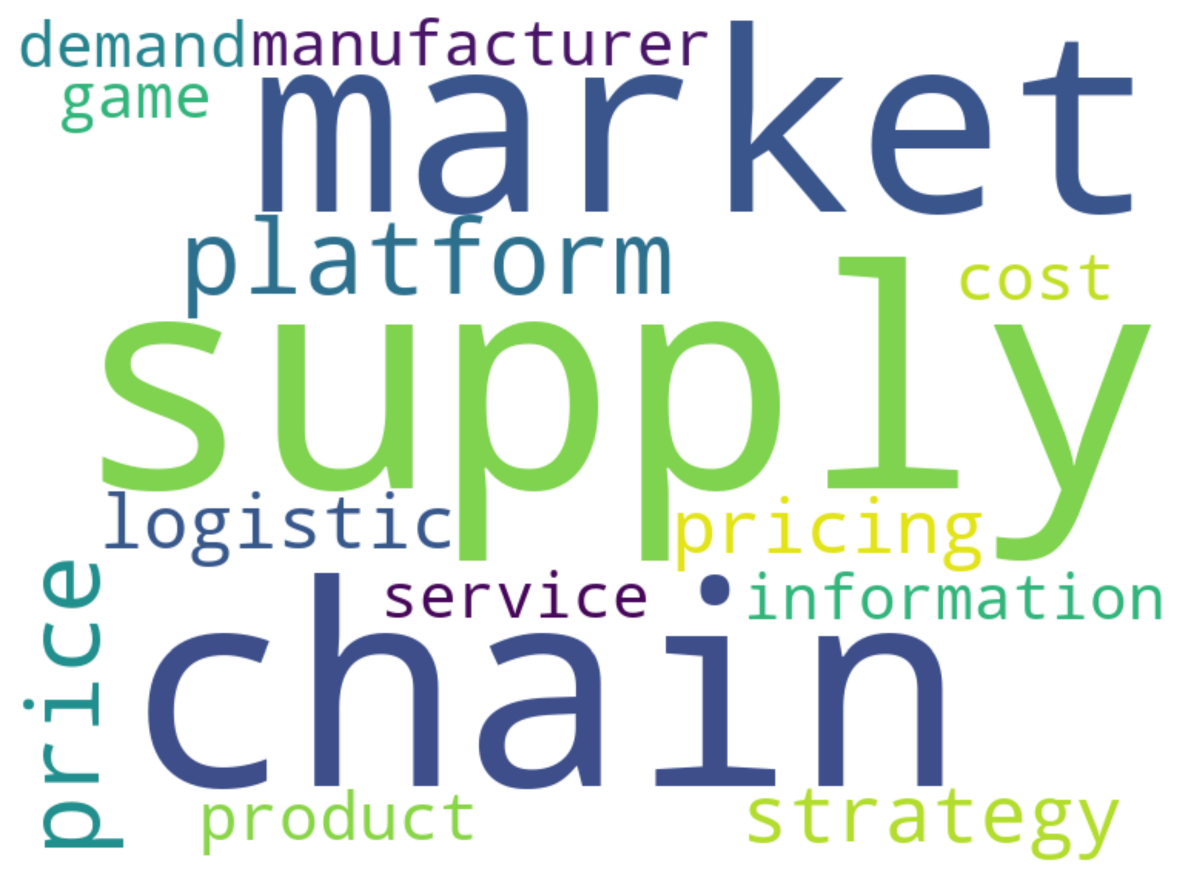}
        \caption{Supply chain}
        \label{fig:tbd}
    \end{subfigure}
    \hfill
    \begin{subfigure}[t]{0.32\textwidth}
        \centering
        \includegraphics[width=0.8\textwidth]{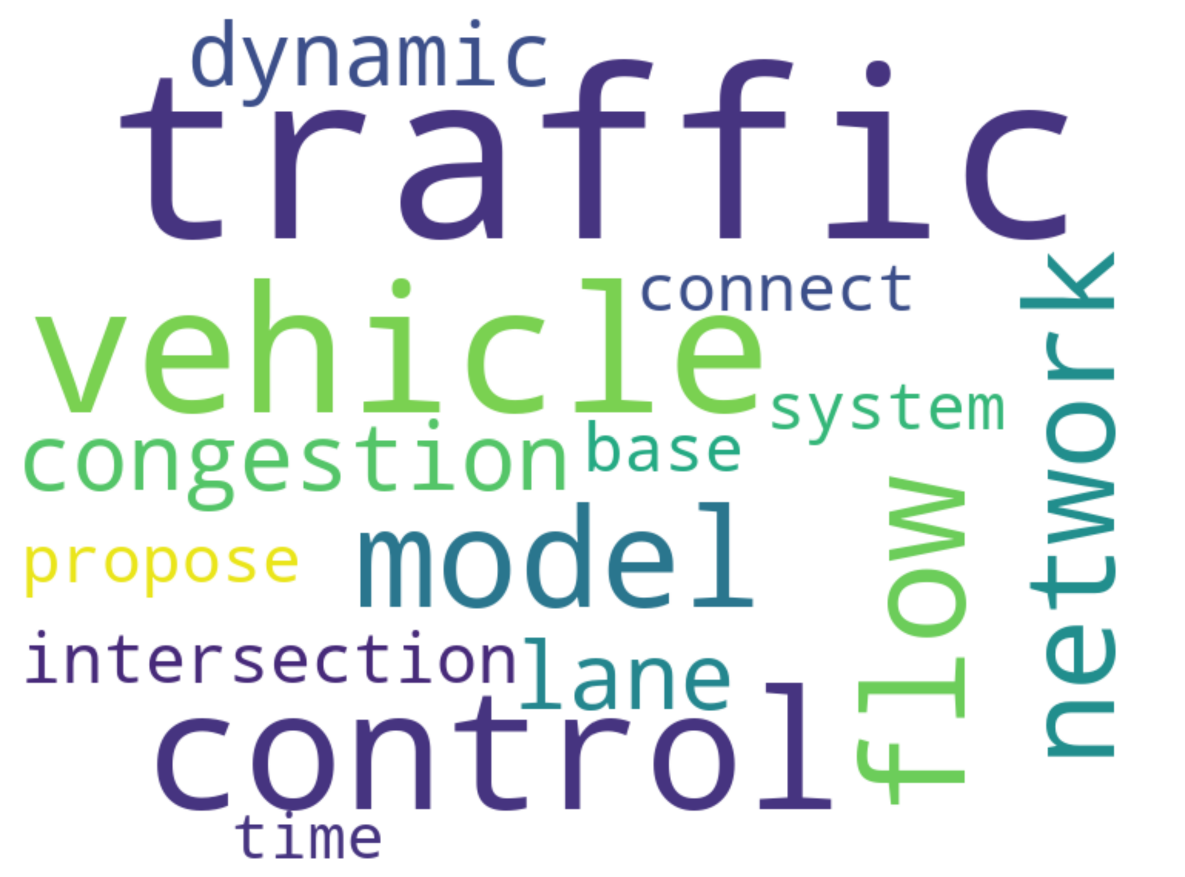}
        \caption{Traffic flow and control}
        \label{fig:traffic-flow}
    \end{subfigure}
    \hfill
    \begin{subfigure}[t]{0.32\textwidth}
        \centering
        \includegraphics[width=0.8\textwidth]{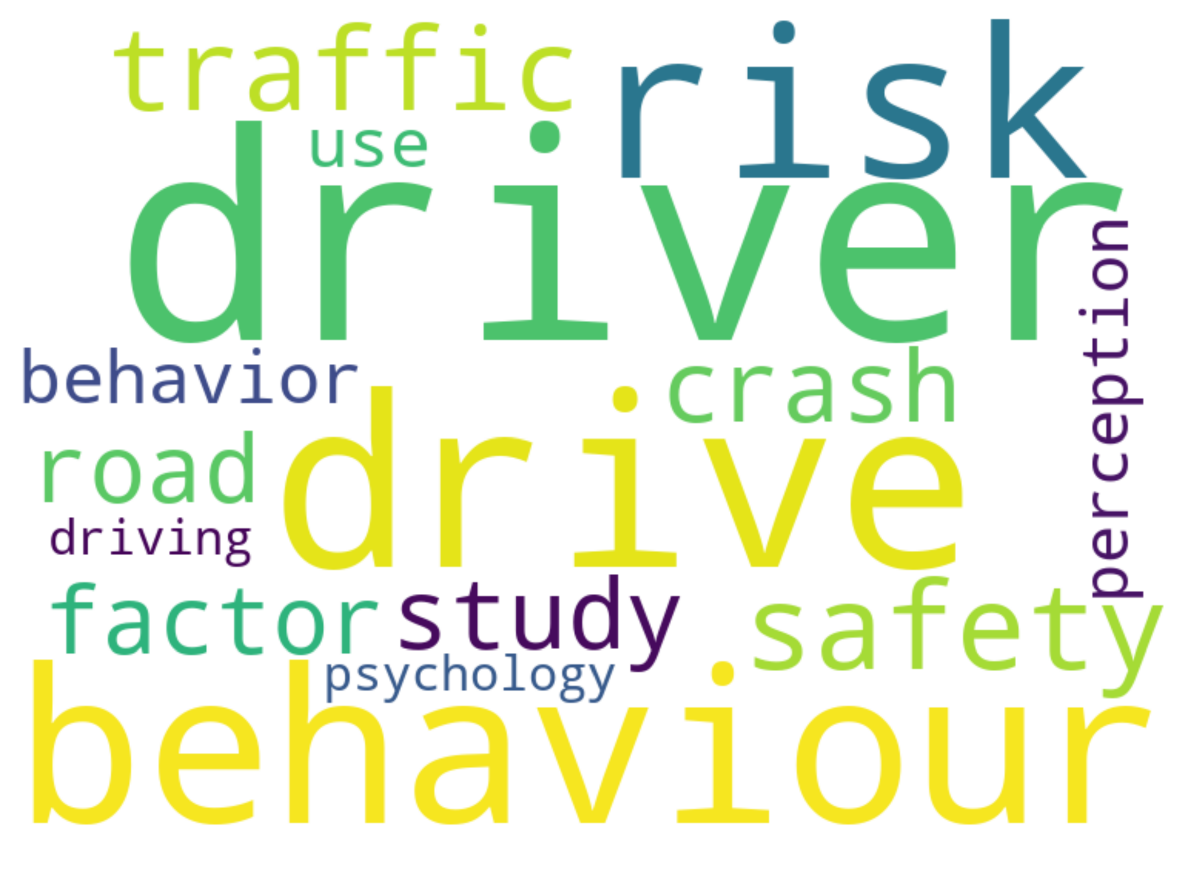}
        \caption{Safety aspects of driving behavior}
        \label{fig:emissions-policy}
    \end{subfigure}

    \vspace{0.5em}

    \begin{subfigure}[t]{0.32\textwidth}
        \centering
        \includegraphics[width=0.8\textwidth]{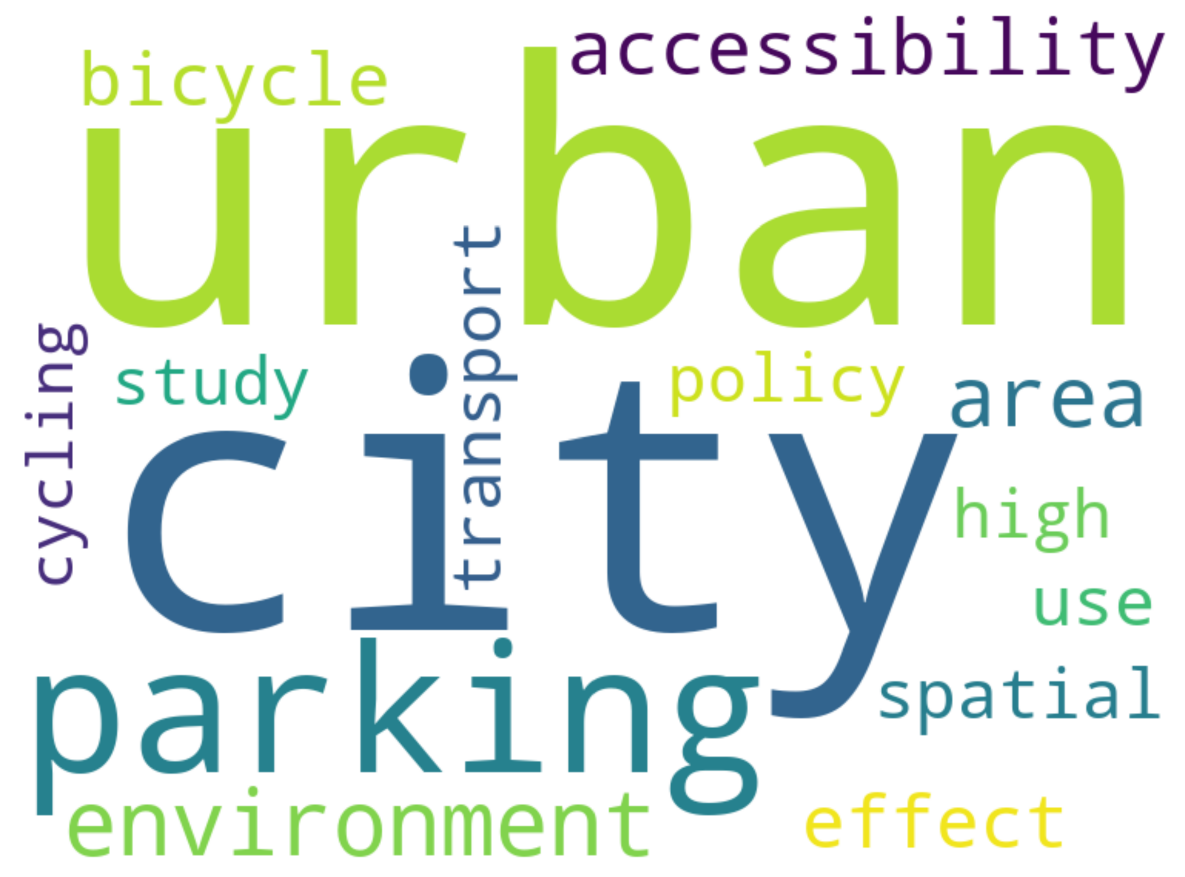}
        \caption{Active transportation and land use}
        \label{fig:mode-choice}
    \end{subfigure}
    \hfill
    \begin{subfigure}[t]{0.32\textwidth}
        \centering
        \includegraphics[width=0.8\textwidth]{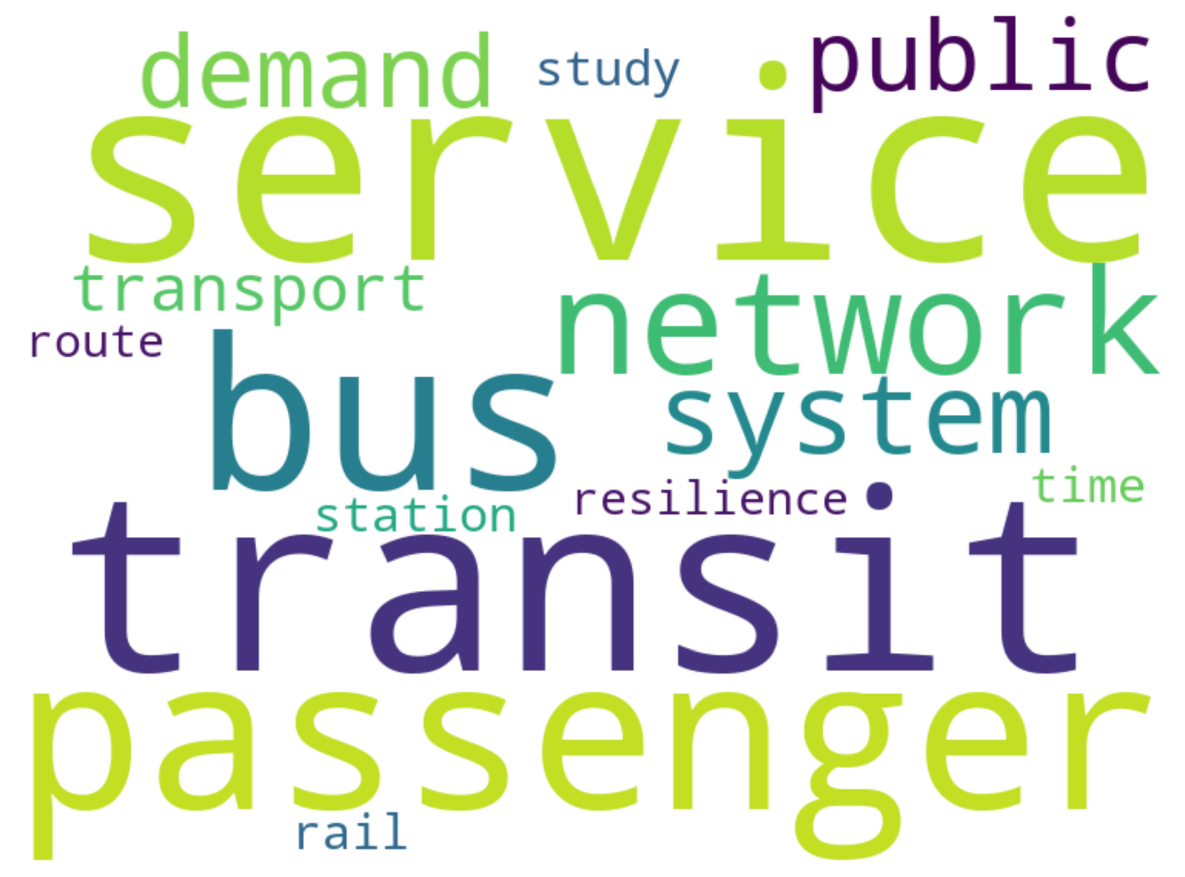}
        \caption{Public transit}
        \label{fig:urban-env}
    \end{subfigure}
    \hfill
    \begin{subfigure}[t]{0.32\textwidth}
        \centering
        \includegraphics[width=0.8\textwidth]{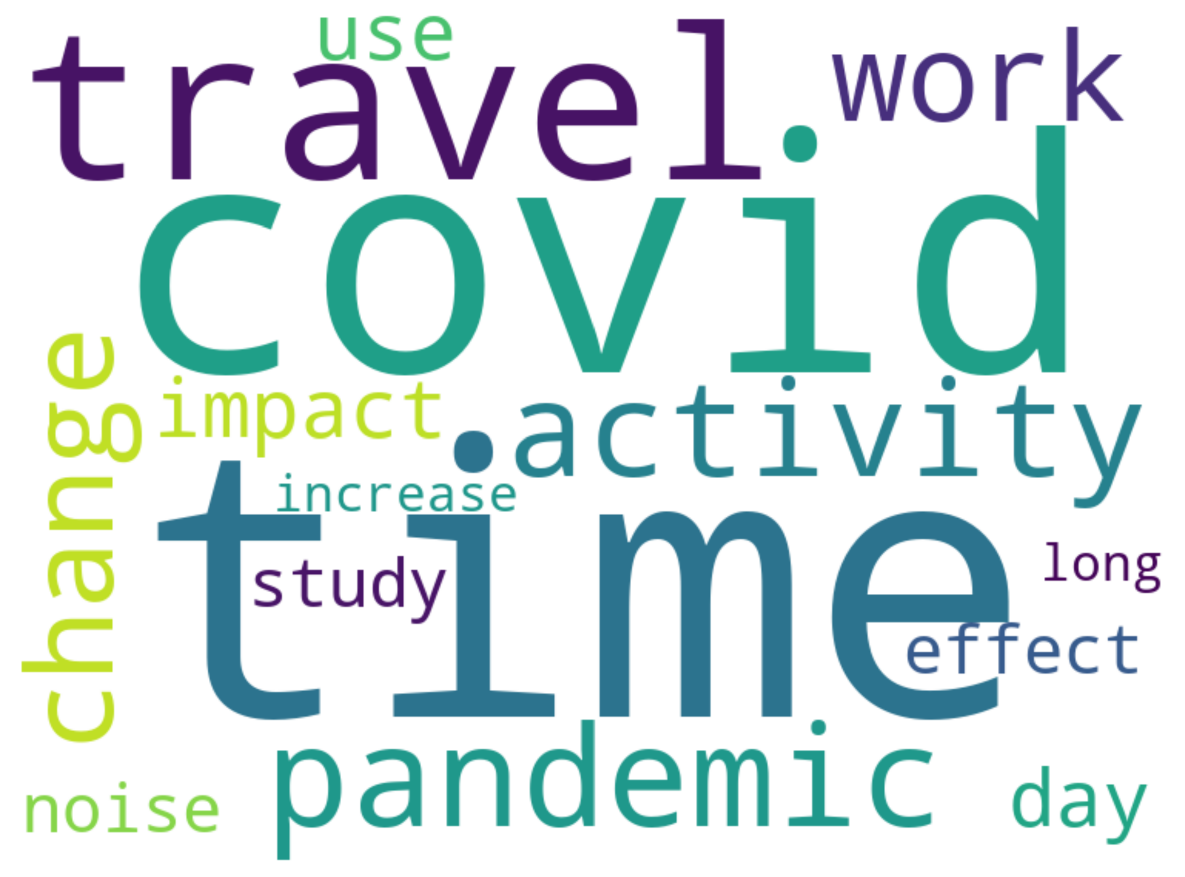}
        \caption{COVID-19 and pandemic}
    \end{subfigure}

    \vspace{0.5em}
    \begin{subfigure}[t]{0.32\textwidth}
        \centering
        \includegraphics[width=0.8\textwidth]{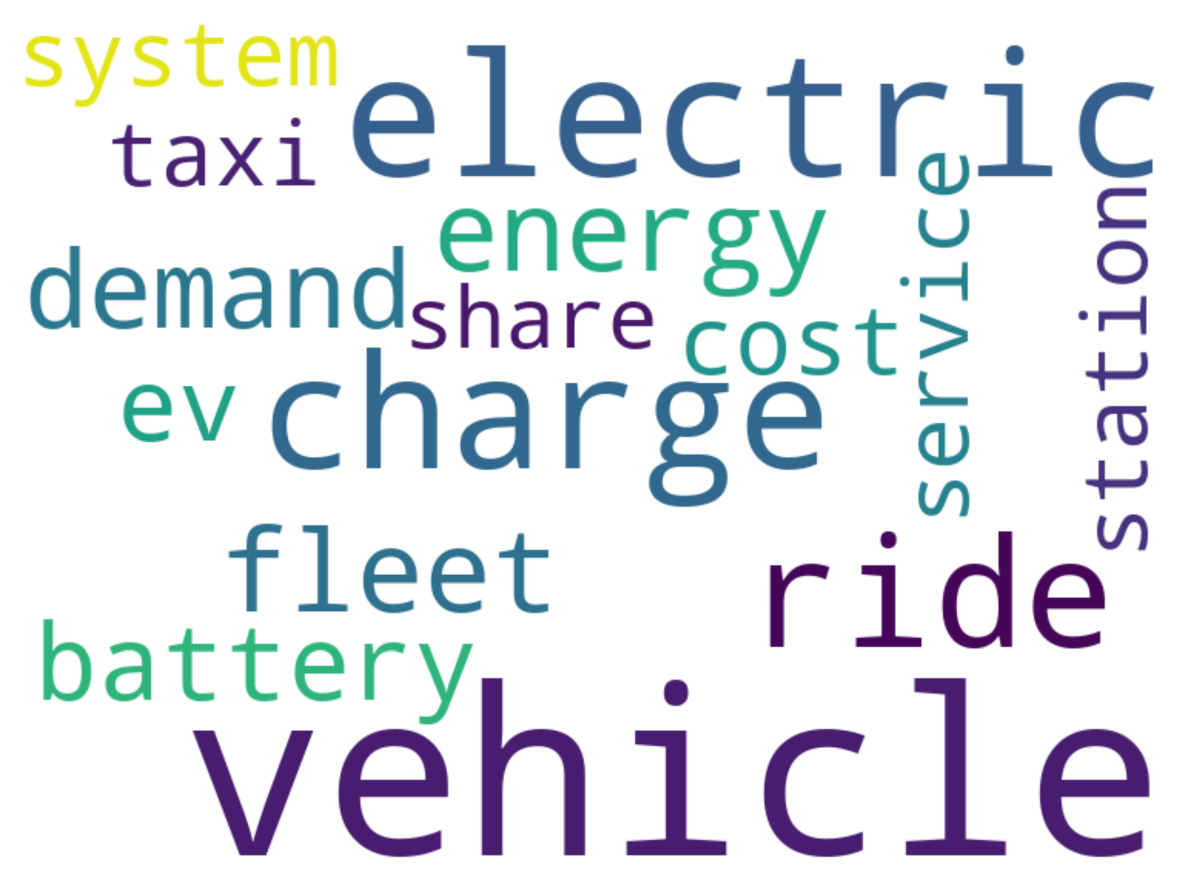}
        \caption{Electric vehicles and ride-sharing}
    \end{subfigure}
    \hfill
    \begin{subfigure}[t]{0.32\textwidth}
        \centering
        \includegraphics[width=0.8\textwidth]{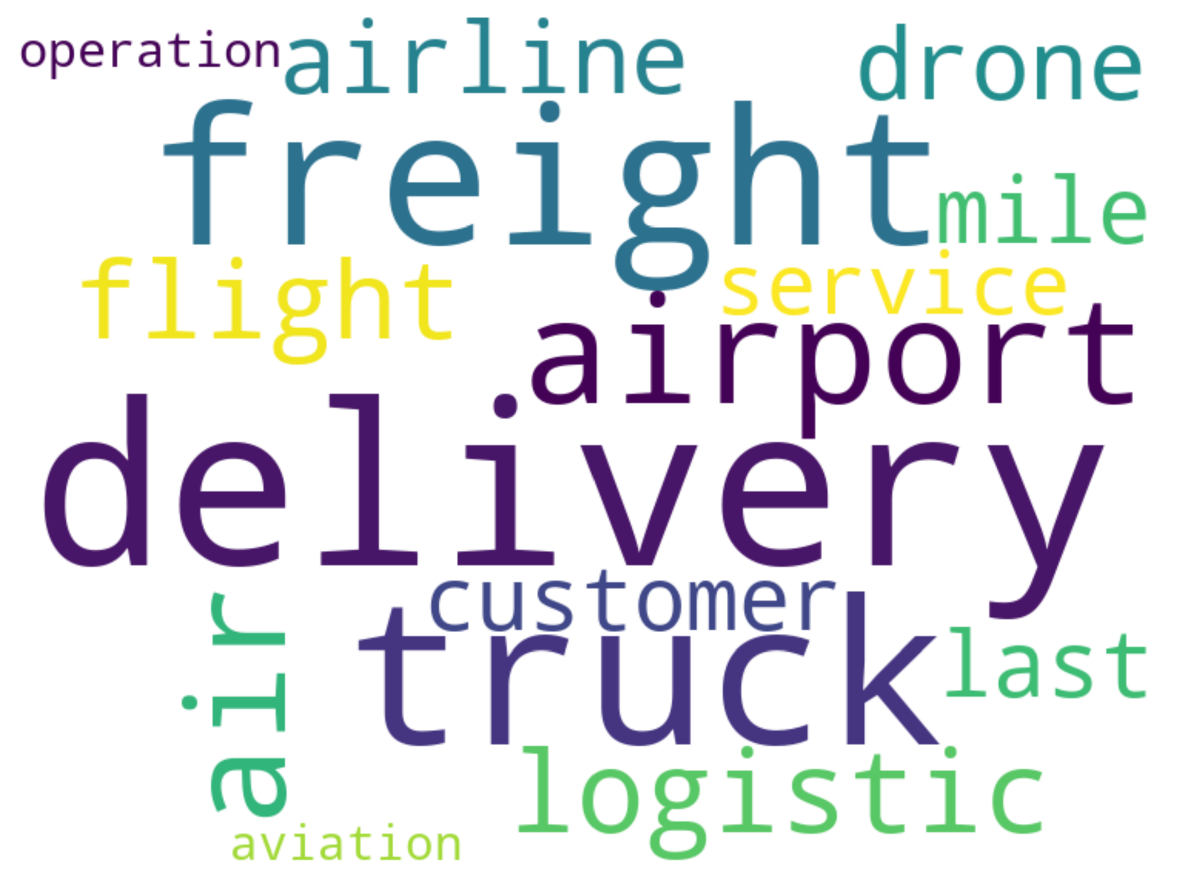}
        \caption{Air and freight logistics
        }
    \end{subfigure}
    \hfill
    \begin{subfigure}[t]{0.32\textwidth}
        \centering
        \includegraphics[width=0.8\textwidth]{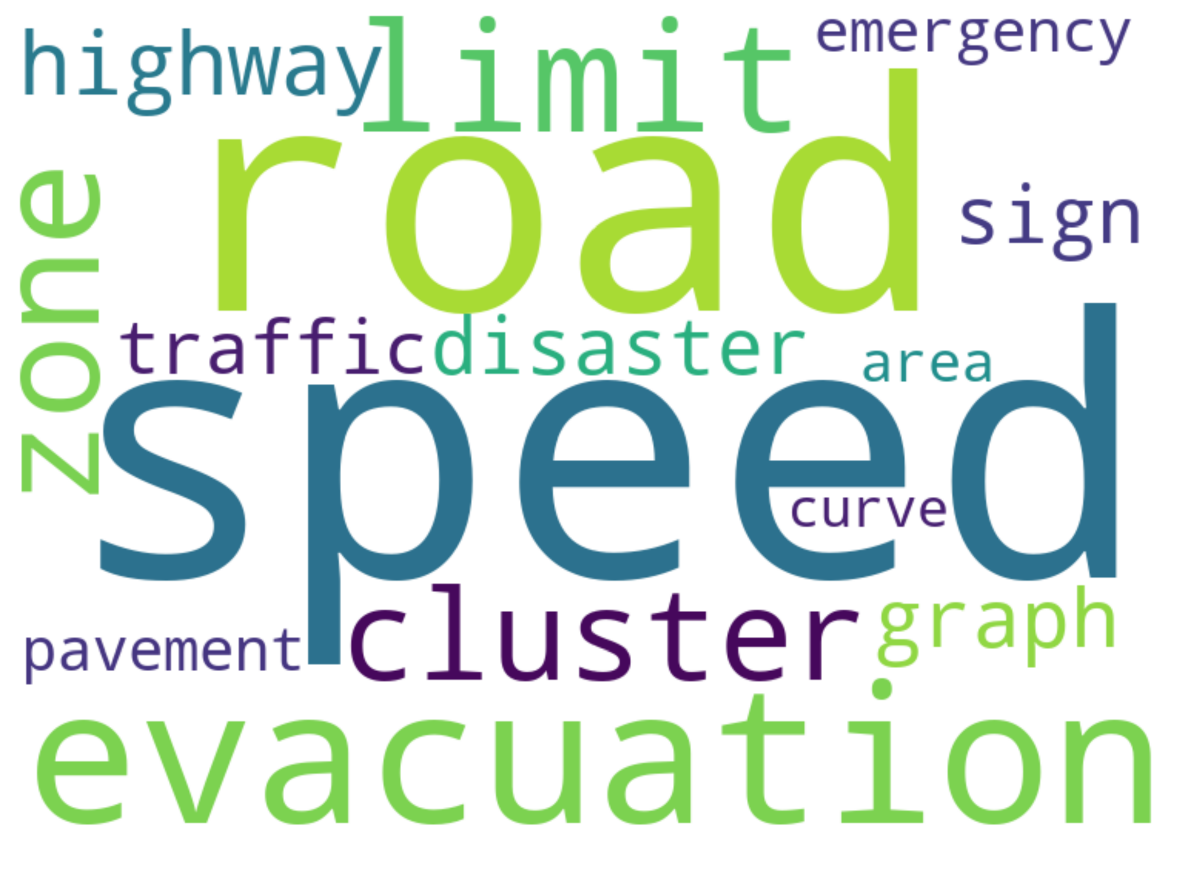}
        \caption{Resilience and evacuation}
    \end{subfigure}
    \caption{Word cloud visualization for each identified topic.}
    \label{fig:main-grid}
\end{figure}

\section{Other models and prompts}
\label{appendix:diff-model}

The exact same pipeline was used with a number of other models, some with more parameters, and some with thinking~\cite{zhong2023chatgptunderstandtoocomparative}.
As shown in Table \ref{tab:model_comparison_wide}, outside of \texttt{is\_code\_publicly\_available}, most of the other models perform within the same interval as the one we used, Gemini-2.5-Flash-Lite. This confirms that our pipeline is not dependent on the exact model used, and can be updated with newer models if needed.

\begin{table*}[ht]
\centering
\caption{Comparison of Agreement Metrics Across LLM Configurations (PA: percent agreemend; FK: Fleiss Kappa)}
\label{tab:model_comparison_wide}
\resizebox{\textwidth}{!}{%
\begin{tabular}{l cc cc cc cc cc}
\toprule
 & \multicolumn{2}{c}{\textbf{\makecell{Gemini 2.5 Flash\\(w/ reason)}}} & \multicolumn{2}{c}{\textbf{Qwen3.6-27B}} & \multicolumn{2}{c}{\textbf{\makecell{GPT\\(reasoning)}}} & \multicolumn{2}{c}{\textbf{\makecell{GPT\\(no reasoning)}}} & \multicolumn{2}{c}{\textbf{\makecell{Gemini 2.5 Flash\\Lite (used in this study)}}} \\
\cmidrule(lr){2-3} \cmidrule(lr){4-5} \cmidrule(lr){6-7} \cmidrule(lr){8-9} \cmidrule(lr){10-11}
\textbf{Variable} & \textbf{PA} & \textbf{FK} & \textbf{PA} & \textbf{FK} & \textbf{PA} & \textbf{FK} & \textbf{PA} & \textbf{FK} & \textbf{PA} & \textbf{FK} \\
\midrule
is\_code\_publicly\_available & 0.9688 & 0.8310 & 0.9583 & 0.7945 & 0.9375 & 0.7273 & 0.9167 & 0.6618 & 0.9688 & 0.8388 \\
is\_data\_repository\_available & 0.8542 & 0.3110 & 0.9062 & 0.4374 & 0.9062 & 0.3243 & 0.8854 & 0.2721 & 0.9167 & 0.3994 \\
is\_quantitative\_study & 0.8125 & 0.3997 & 0.8125 & 0.4146 & 0.8125 & 0.3840 & 0.8750 & 0.3836 & 0.8854 & 0.4586 \\
is\_data\_cited & 0.7292 & 0.5772 & 0.7812 & 0.6783 & 0.7604 & 0.6283 & 0.5833 & 0.4335 & 0.6979 & 0.5224 \\
\bottomrule
\multicolumn{11}{l}{\footnotesize \textbf{Note:} PA = Percentage Agreement, FK = Fleiss Kappa.} \\
\end{tabular}%
}
\end{table*}

To test the effect of changing prompts on the pipeline using Gemini-2.5-Flash-Lite (which does not have thinking enabled), we also used two different prompts to extract results: a prompt that consolidated all features (Oneshot), and one that told the model to manually use chain of thought (Manual CoT).
As expected, Table~\ref{tab:prompt_comparison} shows that oneshot prompts performed worse overall, likely because many of the prompts were not similar, and the model struggled to handle them all at once. On the other hand, using manual CoT only improved results for \texttt{is\_data\_cited}. This may be because Gemini-2.5-Flash-Lite is not powerful enough to derive useful results from CoT, or the CoT prompting was not fine-tuned enough for this specific task.

\begin{table}[htbp]
\centering
\caption{Comparison of Prompting Methods}
\label{tab:prompt_comparison}
\begin{tabular}{lcccccc}
\toprule
 & \multicolumn{2}{c}{\textbf{Manual CoT}} & \multicolumn{2}{c}{\textbf{Oneshot}} & \multicolumn{2}{c}{\textbf{\textit{Multiple Prompts}}} \\
\cmidrule(lr){2-3} \cmidrule(lr){4-5} \cmidrule(lr){6-7}
\textbf{Features} & \textbf{\% Agr.} & \textbf{Fleiss $\kappa$} & \textbf{\% Agr.} & \textbf{Fleiss $\kappa$} & \textbf{\% Agr.} & \textbf{Fleiss $\kappa$} \\
\midrule
\texttt{is\_code\_publicly\_available}      & 0.9688 & 0.8388 & 0.9688 & 0.8388 & 0.9688 & 0.8388 \\
\texttt{is\_data\_repository\_available}  & 0.9167 & 0.4706 & 0.9167 & 0.3043 & 0.9167 & 0.3994 \\
\texttt{is\_quantitative\_study}           & 0.7917 & 0.3495 & 0.8958 & 0.5913 & 0.8854 & 0.4586 \\
\texttt{is\_data\_cited}                  & 0.6979 & 0.5091 & 0.7188 & 0.5397 & 0.6979 & 0.5224 \\

\bottomrule
\end{tabular}
\end{table}

We also test the many-shot in-context learning (ICL)~\cite{agarwal2024many}. To characterize its effect, we report a focused many-shot ICL experiment on the manual validation set. To avoid contaminating the evaluation, we did not place all of our validated ground truth into the prompt, as this would leave no held-out samples for a fair comparison. We instead used 20 of our 96 validated samples with full human agreement as in-context examples and report results on the remaining 76 (Table~\ref{tab:many_shot}).

\begin{table}[htbp]
\centering
\caption{Results of many-shot in-context prompting.}
\label{tab:many_shot}
\begin{tabular}{lcc}
\toprule
 & \multicolumn{2}{c}{\textbf{Results}} \\
\cmidrule(lr){2-3}
\textbf{Features} & \textbf{\% Agr.} & \textbf{Fleiss $\kappa$} \\
\midrule
\texttt{is\_code\_publicly\_available}      & 0.7188 & 0.1219 \\
\texttt{is\_data\_repository\_available}  & 0.6979 & -0.0195 \\
\texttt{is\_quantitative\_study}           & 0.8646 & 0.3015 \\
\texttt{is\_data\_cited}                  & 0.6042 & 0.2190 \\
\bottomrule
\end{tabular}
\end{table}

As shown in Table~\ref{tab:many_shot}, performance under this configuration is lower than under our other prompting methods, particularly for \texttt{is\_code\_publicly\_available} and \texttt{is\_data\_repository\_available}. We do not read this as evidence that many-shot ICL is ineffective in general; rather, three factors specific to our setup likely contribute. First, only 20 examples in this study were available for the prompt, a small set from which to generalize. Second, including these examples expanded the prompt from under 20,000 tokens to roughly 300,000 tokens --- well within the 1M-token limit of Gemini-2.5-flash-lite, but long enough that the result is consistent with the long-context degradation reported for models in this class. Finally, because we used human samples with full agreement in the prompt context, the remaining papers will have a lower agreement. Our design does not fully separate these effects, so we treat the result as empirical rather than conclusive. A stronger frontier model may behave differently, and we note this as a limitation and a direction for future study.

\bibliographystyle{ieeetr}
\bibliography{main.bib}
\end{document}